\documentclass{llncs}

\usepackage{amssymb}
\usepackage{amsfonts}
\usepackage{amsmath}
\usepackage{graphicx}
\usepackage{bussproofs}
\usepackage{verbatim}
\usepackage{color}
\usepackage{xspace}
\usepackage{floatflt}
\usepackage{float}
\usepackage{hyperref}
\usepackage{euscript}
\usepackage{xspace}
\usepackage{euscript}
\usepackage{textcomp}
\usepackage{bbding}
\usepackage{enumerate}
\usepackage{ifthen}
\usepackage{wrapfig}

\usepackage{tikz}
\usetikzlibrary{automata,positioning}

  \newtheorem{fact}[theorem]{Fact}

\def\Pos{\ensuremath{\text{\rm Pos}}}

\newcommand \match[2]{#1 \ll #2}
\newcommand \matchq[2]{#1 \stackrel{?}{\ll} #2}
\newcommand \AbelardM[1]{\big( #1 \big)_{\texttt{A}}}
\newcommand \EloiseM[2]{\big( #1, #2 \big)_{\texttt{E}}}

\newcommand \red[1]{\textcolor{red}{#1}}
\newcommand  \blue[1]{\textcolor{blue}{#1}}

\newcommand \dual[1]{\widetilde{#1}}

\newcommand \uberTo[1]{\stackrel{#1}{\rightarrow}}

\newcommand \bs[1] {\boldsymbol{#1}}

\newcommand \ucomment[1]{}
\newcommand \unnecessary[1]{}

\newcommand \set[1]{\{{#1}\}}

\newcommand \citep[1]{\cite{#1}}

\newcommand \LongVersion[1]{}

\newcommand \tand {\textrm{ and }}
\newcommand \tor {\text{ or }}
\newcommand \model[1]{\langle #1 \rangle}
\newcommand \ttt[1]{\texttt{#1}}

\newcommand \terms {\ensuremath{\mathcal{T}(\Sigma,\mathcal{X})}}
\newcommand \mterms{\ensuremath{\overline{\mathcal{T}}(\Sigma,\mathcal{X})}}
\newcommand \mcons{\ensuremath{\Sigma \cup \mathcal{X}}}
\newcommand \ccons{\ensuremath{(\Sigma \cup \mathcal{X})^{!?}}}

\newcommand {\tst}{ \textrm{ s.t. } }

\newcommand \Abelard {\texttt{Abelard}\xspace}
\newcommand \Eloise {\texttt{Eloise}\xspace}
\newcommand {\vaut}[1] 
   {
  \ifthenelse{\equal{#1}{automaton}}{FVA\xspace}{}
  \ifthenelse{\equal{#1}{automata}}{FVAs}{}
  }
\newcommand {\gsim}[1] {$\EuScript{G}$-{#1}}
\newcommand {\mysim}{\preceq}

\newcommand {\Eu}[1]{\EuScript{#1}}
\newcommand {\eps} {\varepsilon}

\newcommand{\mycal}[1]{\mathcal{#1}}
\newcommand{\rTo}{\longrightarrow}
\newcommand \pfun {\rightharpoonup} 

\newcommand{\synch}{\Join}

\newcommand{\gvert}{\;\;|\;\;}

\newcounter{casecounter}
\def\thecasecounter{(\roman{casecounter})}
\newenvironment{proofbycases}[1][0]{
  \setcounter{casecounter}{#1}%
}{}

\newenvironment{caseinproof}{%
  \refstepcounter{casecounter}%
  \vskip 1pt
  \noindent%
  \blue{\emph{Case~\thecasecounter.}}~
}{\par}

\title{Fresh-Variable Automata for Service Composition}
\begin{document}
\author{Walid Belkhir \inst{1} and Yannick Chevalier \inst{2} and Michael Rusinowitch \inst{1}}

\institute{%
INRIA Nancy--Grand Est \& LORIA  \ 
\email{walid.belkhir@inria.fr,rusi@loria.fr}
\and
Universit\'e Paul Sabatier \&  IRIT Toulouse \ 
\email{ychevali@irit.fr}
}

\maketitle

\begin{abstract}
To model Web services 
handling  data from an infinite domain,  or with multiple sessions, 
we introduce \emph{fresh-variable automata}, a simple extension of  finite-state automata  
in which some transitions are labeled with variables  
that can be refreshed in some specified states. 
We  prove several closure properties for this class of automata 
and study their decision problems. 
We then introduce a notion of simulation that enables us to reduce
the Web service composition problem to the construction of a simulation
of a target service by the asynchronous product of existing services, and prove
that this construction is computable.
 \end{abstract}



\section{Introduction}

Service Oriented Architectures (SOA) consider services as platform-independant elementary components 
that can be published,  invoked over a network and  loosely-coupled with other 
services through standardized XML protocols 
in order to  dynamically build complex  distributed applications~\cite{Alonso04}. 
This flexible ability to compose applications can be viewed as a motto for SOA. 

Service composition has been adressed in many works (e.g.
\cite{Pistore:2005:ACW,icsoc/BerardiCGLM03,colombo,HassenNT08,SCC-WSCA08,BasuBO12}).
One of the most successful approaches to composition amounts to
abstract services as finite-state automata (FA) and apply available
tools from automata theory to synthesize a new service satisfying the
given client requests from an existing community of
services~\cite{icsoc/BerardiCGLM03,colombo,MW08}.

However FA models are too abstract for handling data values ranging
over unbounded domains, such as integer parameters of procedures or
strings attached to XML documents leaves. This limitation has
motivated several extensions of automata for dealing with infinite
alphabets. A noticeable one is \emph{finite-memory automata} (FMA)
proposed by Kaminski and Francez \cite{tcs/KaminskiF94}, studied
and compared with \emph{pebble automata} in~\cite{NevenSV04}.  FMA
have been extended to \emph{data automata}
(e.g.~\cite{Bouyer02,BojanczykLICS06,Segoufin06}) that have better connections with
logic while keeping good decidability properties.  Basically FMA can
only remember a bounded number of previously read symbols. For
instance, they can recognize the language of words where some data
value occurs an even number of times.  Our work is related to
\emph{variable  automata} a simple extension of FA introduced by
\cite{GKS10}.  In this approach some automata transitions are labelled
by variables that can get values from an infinite alphabet. The model 
in \cite{GKS10} allows one to keep a natural definition for runs and 
to obtain simple procedures for membership and non-emptyness.

However  it is not obvious whether   the automata-based  approach to service composition 
(e.g. \cite{icsoc/BerardiCGLM03,MW08}) can still be applied
with infinite alphabets. 
Our objective is to  define  a  class of automata on infinite alphabets 
which is well-adapted to specification and composition of services  and to study its properties. 

\paragraph{Contributions.}
In this paper we consider  the service composition problem 
as stated  in \cite{BerardiCGP08}: \emph{given a client and a community of 
available services, synthesize a composition, i.e. a suitable function 
that delegates actions requested by the client to the available services 
in the community}. This problem amounts to show (\cite{BerardiCGP08,MW08}) that there exists a simulation relation 
between the target service (specifying an expected service behaviour for satisfying the client requests) 
and the asynchronous product 
of the available services. If a simulation relation exists then it can be easily used to 
generate an orchestrator, that is a function that selects  at each step an available 
service for executing an action requested by the client. 
In order to head for real-world applications where service actions are parameterized by  terms 
built with data taken from infinite alphabets (identifiers, codes, addresses \ldots), 
we introduce an extension of FA called \emph{Fresh-Variable Automata} (FVA) where some transitions are labelled by 
variables that can be assigned the read letter.  
A variable binding can be released at some states: in that case we say that 
the variable is \emph{refreshed}.
This mecanism is natural to express  iteration processes, for instance when 
a service has to scan a list of item identifiers, or sessions.  
Note that our freshness notion differs from the one in \cite{Tzevelekos:2011:FA}. 
%
We have established   closure properties of FVA for union, intersection, concatenation and Kleene operator.
We have shown that universality is decidable. 
Our main result is the decidability of  the service composition problem for FVA. 
This gives  a non-trivial extension of 
\cite{BerardiCGP08} that we  illustrate with a natural example.

\paragraph{Related work.}
The  related formalism of variable automata \cite{GKS10} 
was proposed as another simple extension of FA to infinite  alphabets. 
The variables of variable automata are assigned at most once a value in a run,  
except for a special free variable  that can get a value
that is different from  the other variables. This is not convenient 
to model services where several variables are reused in each session.  
\cite{GKS10}  investigates closure properties of variable automata 
but do not consider simulation relations. In fact, FVAs and variable  automata are incomparable.
A  well established model to handle infinite alphabets is FMA \cite{tcs/KaminskiF94}. 
Although our model is less   expressive than FMAs, we believe that FVAs  are simpler to handle and to 
visualize, and they  enjoy more decidable properties such as universality.

Several works deal with the problems of service composition and orchestration 
in different settings. 
In the Roman model \cite{icsoc/BerardiCGLM03},  service composition  
was considered where the services are finite automata with no access to data.
A  logic-based approach was devised in \cite{Pathak06modelingweb} to 
solve this problem where the agents  have  access to infinite data.
The client  and the services exhibit infinite-state behavior: 
the transitions are labeled with guards over an infinite    domain.  
In \cite{BalbianiCF09} the communication actions are performed through channels.
Guards/conditions and constraints  on the transitions have been  introduced as well, e.g. \cite{Pathak06modelingweb}. 
Orchestration was studied  in \cite{SCC-WSCA08} for  services with linear behavior in 
presence of security constraints and  where 
the communication actions are arbitrary terms over a given signature.  

\paragraph{Paper organisation.}
Sec. \ref{prelim:sec} recalls standard notions. 
Sec. \ref{vaut:sec} introduces the new  class of  \vaut{automata}.
Sec. \ref{properties:FVA:sec} studies FVAs,
and  shows in particular closure properties and decidability of universality. 
Sec. \ref{simulation:sec}  defines \emph{communicating FVAs}, or CFVAs for short. 
and  introduces the  notion of \gsim{simulation}. 
Sec. \ref{termination:sec} shows that \gsim{simulation} is  decidable for CFVAs.
Sec. \ref{service:synthesis}  applies the results to  service synthesis problems. 
Final remarks and future works are given in Sec. \ref{conclusion:sec}.     

\section{Preliminaries}
\label{prelim:sec}

Let $\mathcal{X}$ be a finite set
of variables, $\Sigma$ an infinite alphabet of letters.
A substitution is an idempotent mapping $\{x_1\mapsto
\alpha_1,\ldots,x_n\mapsto \alpha_n\}\cup \bigcup_{a\in\Sigma}\set{a
  \mapsto a }$ with variables $x_1,
\ldots, x_n$ in $\mathcal{X}$ and $\alpha_1, \ldots, \alpha_n$ in
$\mycal{X} \cup \Sigma $. We call $\set{x_1,\ldots,x_n}$ its
\emph{proper domain}, and denote it by $dom(\sigma)$. We denote by $Dom(\sigma)$ 
the set $dom(\sigma) \cup \Sigma$. 
If all the $\alpha_i,i=1\ldots
n$ are letters then we say that $\sigma$ is ground.  
The empty substitution (\textit{i.e.}, with an empty proper domain) is denoted
by $\emptyset$.  The set of the substitutions 
from $\mycal{X}\cup\Sigma$ to a
set $A$ is denoted by $\zeta_{\mycal{X},A}$, or by $\zeta_{\mycal{X}}$, or simply  by $\zeta$ if
there is no ambiguity.  If $\sigma_1$ and $\sigma_2$ are substitutions
that coincide on the domain $dom(\sigma_1)\cap dom(\sigma_2)$, then
$\sigma_1 \cup \sigma_2$ denotes their union in the usual sense.  We
define the function $\mycal{V}:\Sigma\cup \mycal{X} \rTo
\mycal{P}(\mycal{X})$ by $\mycal{V}(\alpha)=\set{\alpha}$ if $\alpha
\in \mycal{X}$, and $\mycal{V}(\alpha)=\emptyset$, otherwise.
For a function $F: A \rTo B$, and $A'\subseteq A$, the restriction of
$F$ on $A'$ is denoted by $F_{|A'}$. 

A two-players game is a tuple $\langle \Pos_{E},\Pos_{A},M, p^{\star}
\rangle$, where $\Pos_{E},\Pos_{A}$ are  disjoint sets of
positions: \Eloise's positions and \Abelard's positions. $M\subseteq
(\Pos_{E} \cup \Pos_{A}) \times (\Pos_{E} \cup \Pos_{A})$ is a set of
\emph{moves}, and $p^{\star}$ is the starting position.  A strategy
for the player \Eloise is a function $\rho: \Pos_E \uberTo{} \Pos_E \cup
Pos_A$, such that $(\wp,\rho(\wp)) \in M$ for all $\wp \in \Pos_E$. A
(possibly infinite) play $\pi=\langle \wp_1,\wp_2,\ldots \rangle$
\emph{follows} a strategy $\rho$ for player \Eloise iff
$\wp_{i+1}=\rho(\wp_i)$ for all $i \in \mathbb{N}$ such that $\wp_i
\in \Pos_{E}$. Let $\Eu{W}$ be a (possibly infinite) set of plays. 
A strategy $\rho$ is \emph{winning} for \Eloise from a
set $S \subseteq \Pos_E\cup \Pos_A$ according to   $\Eu{W}$
iff every play starting from a
position in $S$ and following $\rho$ belongs to $\Eu{W}$.


\section{Fresh-variable automata}
\label{vaut:sec}
In this section we introduce the class of FVAs and
illustrate it through simple examples.  This formalism 
extends finite-state automata with two features. Firstly,
the transitions labels  consist of letters and variables that can be assigned a value from an infinite alphabet domain.
 Secondly, at each state some of the variables are 
freed from their assignments: they can receive other  values.

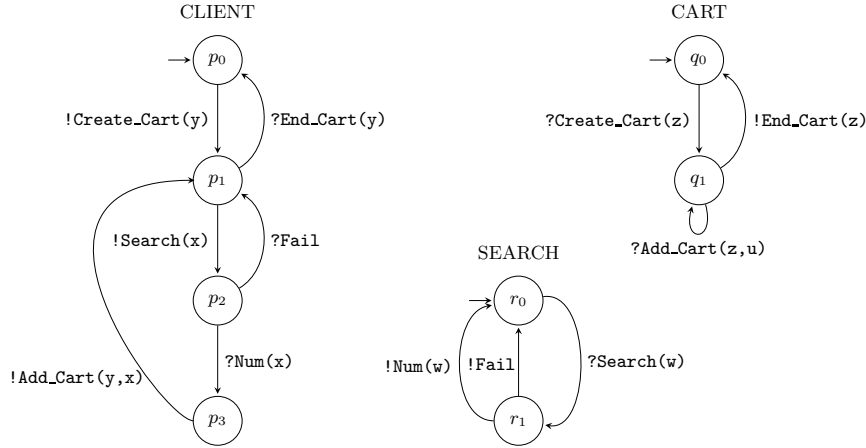
\begin{figure}[h*]
\label{main:example:fig}
  \vspace{-22pt}
\begin{center}
\scalebox{0.8}{\begin{tikzpicture}[shorten >=1pt,node distance=2cm, bend angle=60,
   on grid,auto, initial text=, >=stealth] 
   \node[state,initial] (p_0)   {$p_0$}; 
   \node[state] (p_1) [below =of p_0] {$p_1$}; 
   \node[state](p_2) [below =of p_1] {$p_2$};
   \node[state](p_3) [below =of p_2] {$p_3$};
    \path[->] 
    (p_0) edge  node [left] { \ttt{!Create\_Cart(y)}  } (p_1)
    (p_1) edge   node  [left] {\ttt{!Search(x)}} (p_2)
    (p_2) edge  [bend right] node [right] {\ttt{?Fail}}  (p_1)
    (p_1) edge [bend right]  node [right] { \ttt{?End\_Cart(y)}} (p_0)
    (p_2) edge  node [right] {\ttt{?Num(x)}} (p_3)
    -- (0,0.8) node {CLIENT}
    -- (-2.34,-5.3) node {\ttt{!Add\_Cart(y,x)}};
     \draw[->] (-0.40,-6) .. controls +(left:0.5cm) and +(left:3.5cm) .. (-0.35,-2);    

  \begin{scope}[xshift=8cm, yshift=0cm,node distance=2cm]
   \node[state,initial] (q_0)   {$q_0$}; 
   \node[state] (q_1) [below =of q_0] {$q_1$}; 
    \path[->] 
    (q_0) edge   node [left] { \ttt{?Create\_Cart(z)}} (q_1)
    (q_1) edge  [loop below]  node  {\ttt{?Add\_Cart(z,u)}} ()
    (q_1) edge  [bend right] node [right] {\ttt{!End\_Cart(z)}} (q_0)
    -- (0,0.8) node {CART};
  \end{scope}

  \begin{scope}[xshift=5cm, yshift=-4cm,node distance=2cm,bend angle=99]
   \node[state,initial] (r_0)   {$r_0$}; 
   \node[state] (r_1) [below =of r_0] {$r_1$}; 
    \path[->] 
    (r_0) edge  [bend left]  node [right] {\ttt{?Search(w)}} (r_1)
    (r_1) edge  [bend left,out=90,in=100]  node  [left] {\ttt{!Num(w)}} (r_0)
    (r_1) edge   node [left] {{}} (r_0)
    -- (0,0.8) node {SEARCH}
    -- (-0.46,-1.06) node {\ttt{!Fail}};
  \end{scope}
\end{tikzpicture}}
\end{center}
  \vspace{-22pt}
\caption{The CART  example}
  \vspace{-10pt}
\end{figure}

\paragraph{A motivating example.}
We first motivate fresh-variable automata  through an example that illustrates
 a service composition problem.
 We have an e-commerce  Web site allowing customers 
 to create  shopping carts, search for  items  from an infinite domain and  add them  to 
   a shopping cart, see Figure \ref{main:example:fig}. 
 The main issue is that
 the three agents  CLIENT,  CART and SEARCH  exhibit an infinite-state behavior involving sending and receiving 
  messages ranging over a possibly  infinite set of terms. 
We emphasize that  variable $y$ is refreshed (i.e. freed to get a new value)
when  passing through the state $p_0$. In the same way  variable
 $x$ is  refreshed at  $p_1$,   
 $z$  at $q_0$ and $u$ at $q_1$, and $w$  at $r_0$  respectively.

In this example, we ask whether the   requests  made by the client 
 can be answered by combining the services \ttt{CART} and \ttt{SEARCH}.
In this section we consider only automata in which  the 
transitions are labeled by letters or variables.
We  introduce the communication symbols $!,?$ for defining a  simulation 
in Sec.~\ref{simulation:sec}.


\begin{definition}
  A \emph{FVA} is a tuple
  $A=\model{\Sigma,\mathcal{X},Q,Q_0,\delta,F,\kappa}$ where $\Sigma$
  is a infinite set of letters, $\mathcal{X}$ is a finite set of
  variables, $Q$ is a finite set of states, $Q_0\subseteq Q$ is a set
  of initial states, $\delta=Q \times (\mcons) \to 2^{Q}$ is a
  transition function with finite domain, $F\subseteq Q$ is a set of accepting states,
  and $\kappa: \mycal{X} \rightarrow 2^Q$ is the refreshing function
  that associates to every variable the (possibly empty) set of states
  where it is refreshed.
\end{definition}
For a FVA $\mycal{A}$, we shall denote by $\Sigma_{\mycal{A}}$ the
finite set of letters that appear in the transition function of
$\mycal{A}$. Variables in a FVA are considered up to renaming, and we
always assume that two FVAs have disjoint sets of variables.


\noindent The formal definition of configuration, run and recognized language follows.
\begin{definition}
\label{run:lang:def}
Let   $\mycal{A}=\model{\Sigma,\mathcal{X},Q,Q_0,\delta,F,\kappa}$ be a FVA.  
A \emph{configuration}  is a pair  $(q,M)$ where $q \in Q$ and 
$M:\mycal{X} \pfun  \Sigma$  is a substitution.
 We define a transition relation  over the  configurations 
 as follows:  $(q_1,M_1)\uberTo{a} (q_2,M_2)$, where $a \in \Sigma$,
   iff there exists a label $\alpha \in \Sigma\cup \mycal{X}$ 
  such that  $q_2\in \delta(q_1,\alpha)$,
and either \\
\emph{(i)} $\alpha \in Dom(M_1)$, $M_1(\alpha)=a$ and $M_2={M_1}_{| D}$, with $D=Dom(M_1) \setminus  \kappa^{-1}(q_2)$ or\\
\emph{(ii)} $\alpha \in (\mycal{X}\setminus Dom(M_1))$ and 
 $M_2=(M_1\cup \set{\alpha \mapsto a})_{| D}$, with $D=(Dom(M_1) \cup  \set{\alpha } )\setminus  \kappa^{-1}(q_2)$.\\
   A finite word $w=w_1w_2\ldots w_n \in \Sigma^*$ is \emph{recognized} by 
$\mycal{A}$ iff  there exists a run
$(q_0,M_{0}) \uberTo{w_1}(q_1,M_1)\uberTo{w_2} \ldots \uberTo{w_n}(q_n,M_n)$, such that 
$M_{0}=\emptyset$, $q_0\in Q_0$ and $q_n \in F$.
 The set of words recognized by $\mycal{A}$ is denoted by
$L(\mycal{A})$.
\end{definition}

\newbox\figone
\savebox\figone{\vtop{\begin{tikzpicture}[shorten >=1pt,node distance=2cm, bend angle=60,
   on grid,auto, initial text=, >=stealth] 
   \node[state,initial,accepting] (r_0)   {$p_0$}; 
   \node[state] (r_1) [right =of r_0] {$p_1$}; 
    \path[->] 
    (r_0) [bend left] edge   node [above] {$x$} (r_1)
    (r_1) [bend left] edge  node  [below] {$x$} (r_0)
    -- (-0.7,0.6) node {$\mycal{A}_1$};
    \begin{scope}[xshift=0cm, yshift=2cm,shorten >=1pt,node distance=2cm, bend angle=60,
      on grid,auto, initial text=, >=stealth] 
      \node[state,initial] (q_0)   {$q_0$}; 
      \node[state] (q_1) [right =of q_0] {$q_1$}; 
      \node[state,accepting](q_2) [right =of q_1] {$q_2$};
      \path[->] 
      (q_0) edge [loop above]  node [above] {z} ()
      (q_0) edge  node   {y} (q_1)
      (q_1) edge  [loop above] node [above] {z}  ()
      (q_1) edge  node [above] {y} (q_2)
      -- (-0.7,0.6) node {$\mycal{A}_2$};
    \end{scope}
\end{tikzpicture}}}
\newbox\exampleone
\savebox\exampleone{\vtop{\hbox to 6cm{\vtop{%
\vspace*{-4.5cm}
\parshape=4 0cm 5cm 0cm 5cm 0cm 5cm 0cm 5cm
We could define FVAs with $\varepsilon$-transitions too. We show in
the Appendix that FVAs with $\varepsilon$-tran\-si\-tions are equivalent
to FVAs.

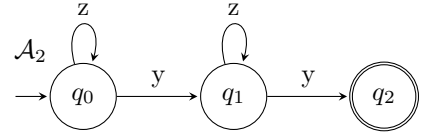
\parshape=8  0cm 5cm 0cm 5cm 0cm 5cm 0cm 5cm 0cm 5cm 0cm 5cm 0cm \textwidth 0cm \textwidth 
\begin{example}
\label{ex:VFA}
Let $\mycal{A}_1$ and $\mycal{A}_2$ be the FVAs depicted on the right,
where $\kappa(x)=\set{p_0}$ and $\kappa(z)=\set{q_0,q_1}$.  Then,
$L(\mycal{A}_1)$ is the set of words $a_1a_1 a_2a_2
\cdots a_na_n$ for $n\ge 0$ and $a_i \in \Sigma$, and
$L(\mycal{A}_2)$ is the set of words in $\Sigma^{\star}$, where some
letter appears at least twice. We notice that $L(\mycal{A}_1)$ cannot
be recognized by a variable automata \cite{GKS10}.
\end{example}}}}}

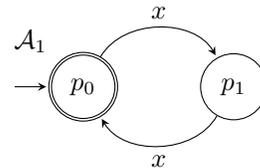
\hbox to \textwidth{\usebox\exampleone\usebox\figone}


\ucomment{\red{[[A DEPLACER]]}
\begin{remark}
Let $\mycal{A}=\model{\Sigma,\mathcal{X},Q,Q_0,\delta,F,\kappa}$ 
and $\mycal{A}'=\model{\Sigma,\mathcal{X},Q,Q_0,\delta,F,\kappa'}$ be two 
FVAs. We notice that if $\kappa \subseteq \kappa'$, then 
$L(\mycal{A}) \subseteq L(\mycal{A}')$.
\end{remark}}


\section{Properties of FVAs}
\label{properties:FVA:sec}
We study in this section properties of FVAs and some basic decision
problems.

\subsection{Closure properties}



\paragraph{FVAs with multiple labels.}
To prove the closure under intersection, we first introduce a
generalization of FVAs called $n$-FVAs where $n$ is an integer. An
$n$-FVA has transitions labeled with $n$-tuple of labels. In this
general setting $1$-FVAs are FVAs. We show next that $n$-FVAs and
FVAs recognize the same languages.

\begin{definition}
  An $n$-\emph{FVA}, where $n \in \mathbb{N}^{\star}$,  is a tuple
  $\mycal{A}=\model{\Sigma,\mathcal{X},Q,Q_0,\delta,F,\kappa}$ which is defined like
   a FVA but for the   transition function  $\delta: Q \times (\Sigma \cup \mycal{X})^n \to 2^{Q}$.
\end{definition}

The configurations and runs of $n$-FVAs are defined as for FVAs,
except that the currently  read letter $u\in \Sigma$  should match simultaneously with the 
$n$ components of its $n$-label for this transition to be fired, see Appendix \ref{NFVA:sec:appendix}.

\begin{theorem}
\label{nFVA:FVA:Th}
For all $n\ge 1$,  $n$-FVAs and FVAs  are equivalent. 
\end{theorem}

\begin{proof}
  We sketch a proof of the non-trivial direction in the case $n=2$.
  The general case follows directly by induction on $n$.  Let
  $\mycal{A}=\model{\Sigma,\mathcal{X},Q,Q_0,\delta,F,\kappa}$ be a
  2-FVA, and let us introduce $n_{\mycal{X}} = \vert \mycal{X} \vert$,
  and $n_{\Sigma} = \vert \Sigma_{\mycal{A}} \vert$, and assume
  $\Sigma_{\mycal{A}}=\set{a_1,\ldots, a_{n_\Sigma}}$.  Let $\Psi
  \subset \set{1,\ldots,n_{\mycal{X}}+n_{\Sigma}}^{\Sigma_{\mycal{A}}
    \cup \mycal{X}}$ be the set of functions from $\Sigma_{\mycal{A}}
  \cup \mycal{X}$ to $\set{1,\ldots,n_{\mycal{X}}+n_{\Sigma}}$ such
  that for every $\psi\in\Psi$ we have $\psi(a_k) = k$. Furthermore,
  given $D\subseteq \mycal{X}$ and $\psi\in\Psi$, we let $\psi^{D}$ be
  the subset of $\Psi$ of functions equal to $\psi$ on
  $(\Sigma_{\mycal{A}} \cup \mycal{X} )\setminus D$. Finally, given a
  substitution $M \in \zeta_{\mycal{X},\Sigma}$ we let $\Psi_{M}$ be
  the subset of $\Psi$ of functions $\psi$ such that, for all $x,y\in
  \Sigma_A\cup dom(M)$, we have $M(x) = M(y)$ \emph{iff} $\psi(x) = \psi(y)$.
  Let $\mycal{A}'=\model{\Sigma,\mycal{X},Q\times\Psi,
    Q_0\times \Psi, F\times\Psi, \delta' , \kappa'} $ where the
  transition function $\delta'$ is defined as follows: for all
  $(q_0,\psi_0) \in Q'$ and $\alpha,\beta \in \Sigma_{\mycal{A}} \cup
  \mycal{X}$, {$\delta'( (q_0,\psi_0) , (\alpha,\beta) ) = \lbrace (
    q_1 , \psi_1) \,\vert\, q_1 \in \delta ( q_0 , (\alpha,\beta))
    \text{ and } \psi_0(\alpha)=\psi_0(\beta) \text{ and } \psi_1 \in
    \psi_0^{\kappa^{-1}(q_1)} \rbrace$}; Finally for $x\in\mycal{X}$,
  we define $\kappa'(x)=\kappa(x)\times\Psi$.
  We can prove that there exists a run $q_0,M_0
  \uberTo{(\alpha_1,\beta_1)}  \ldots
  \uberTo{(\alpha_n,\beta_n)} q_n,M_n$ in $\mycal{A}$ \emph{iff} for
  all $\psi_n \in \Psi_{M_n}$ there exists a run $(q_0,\psi_0),M_0
  \uberTo{(\alpha_1,\beta_1)}  \ldots
  \uberTo{(\alpha_n,\beta_n)} (q_n,\psi_n),M_n$ in $\mycal{A}'$. Thus,
  $\mycal{A}$ and $\mycal{A}'$ recognize the same language $L$.
  Finally, a 1-FVA $\mycal{B}$ recognizing the same language $L$ is
  constructed from $\mycal{A}'$ by mapping each integer in
  $\psi(\mycal{X}\cup\Sigma_{\mycal{A}})$ to a variable or a constant.\qed
\end{proof}

As shown by a language $L=\set{a}$, with $a\in\Sigma$, the
complement of a FVA(-recognizable) language is not necessarily
FVA-recognizable.  Note also that \cite{GKS10} has neither considered
Kleene operator nor the concatenation. The closure under union is
straightforward since we just take the disjoint union of the two FVAs.
The closure under Kleene operation and concatenation is a direct
consequence of the fact that FVAs with $\varepsilon$-transitions and
FVAs are equivalent (Lemma \ref{eps:closed:lemma} in the Appendix).
The closure under intersection is an immediate consequence of
Theorem~\ref{nFVA:FVA:Th}, since the intersection of two FVAs amounts
to computing their Cartesian product, which is a $2$-FVA. Thus we have
the following theorem.
\begin{theorem}
\label{closure:Th}
FVAs are closed under union, concatenation, Kleene operator and
intersection.
\end{theorem}

\subsection{Decision procedures for FVAs}
\label{decision:proc:FVA:sec}
We study the decidability and complexity of classical decision
problems: Nonemptiness (given $\mycal{A}$, is $L(\mycal{A})\neq
\emptyset$?), Membership (given a word $w$ and $\mycal{A}$, is $w \in
L(\mycal{A})$?), Universality (given $\mycal{A}$, is
$L(\mycal{A})=\Sigma^*$?), and Containment (given $\mycal{A}_1$ and
$\mycal{A}_2$, is $L(\mycal{A}_1) \subseteq L(\mycal{A}_2)$?).


\begin{theorem}
\label{univ:FVA:th}
For FVAs, Nonemptiness is NL-complete, Membership is NP-com\-plete, and
Universality is decidable. 
\end{theorem}

\begin{trivlist}
\item {Proof for Universality.} We say a variable $x$ is free in a configuration
  $q,M$ if $x\notin dom(M)$.  Out of $\mycal{A}$ we construct a
  FVA $\mycal{A}'$ such that for every reachable configuration $q',M$ on
  $\mycal{A}'$ every transition out of $q'$ is labeled with a variable
  free in $q'$. 
  \begin{trivlist}
  \item\textbf{Claim 1. }\em
    If $\mycal{A}$ is universal then for every $n\ge 0$ there exists a
    path of length $n$ from an initial state to a final state in which
    every transition is labeled  with a   variable which is free in 
    the  source  state of this transition.
  \end{trivlist}
  \begin{trivlist}
  \item \textsc{Proof of the claim.}  By contradiction assume
    $\mycal{A}$ is universal but there exists $n\ge 0$ such that every
    path of length $n$ from an initial state to a final state has at
    least one transition over either a letter or an already bound
    variable. We note that the word $w_1\ldots w_n \in
    \Sigma^{\star}$, in which $w_i\neq w_j$ for all $i\neq j$ and $w_i
    \notin \Sigma_{\mycal{A}}$, is not recognized by $\mycal{A}$. This
    contradicts the universality of $\mycal{A}$.\qed
  \end{trivlist}
  Assume $\mycal{A}=\model{\Sigma,\mycal{X},Q,Q_0,F,\delta,\kappa}$
  and let
  $\mycal{A}'=\model{\Sigma,\mycal{X},Q',Q_0',F',\delta',\kappa'}$
  where:
  $$
  \left\lbrace
    \begin{array}{rl}
      Q' &= \set{ (q,X) \,\vert\, q\in Q \text{ and }X\subseteq
        \mycal{X}}\\
      Q_0' &= \set{ (q,\mycal{X}) \,\vert\, q\in Q_0 }\\
      F' &= \set{ (q,X) \,\vert\, q\in F \text{ and }X\subseteq
        \mycal{X} }\\
    \end{array}
  \right.
  $$
  and $(q',X')\in\delta'((q,X),x)$ if, and only if, $x\in X$ and
  $X'=(X\setminus\set{x})\cup \kappa^{-1}(q')$, and $\kappa'(x)=
  \set{(q,X) \,\vert\, q\in\kappa(x)}$.  
  \begin{trivlist}
  \item \textbf{Claim 2. }\em 
    There exists a run $q_0,M_0\uberTo{x_1}\ldots\uberTo{x_n} q_n,M_n$
    in $\mycal{A}$ in which for all $1\le i\le n$ we have $x_i\notin
    Dom(M_{i-1})$ if, and only if, there exists a run
    $(q_0,\mycal{X}),M_0\uberTo{x_1}\ldots\uberTo{x_n} (q_n,X_n),M_n$
    in $\mycal{A}'$ with $X_n=\mycal{X}\setminus Dom(M_n)$.
  \end{trivlist}
  \begin{trivlist}
  \item \textsc{Proof of the claim.} By induction on $n$. Since
    $dom(M_0)=\emptyset$ the case $n=0$ is trivial. Assume the claim
    holds up to  $n$. Let us prove the equivalence for
    $n+1$.
    \begin{trivlist}
    \item[$\Leftarrow$)] Since $(q_{n+1},X_{n+1})\in
      \delta'((q_n,X_n),x_{n+1})$ by induction $x_{n+1}\notin
      Dom(M_n)$.  Thus $q_{n+1}\in\delta(q_n,x_{n+1})$ and $x_{n+1}$
      is free at the state $q_n$ of the run. The substitution 
      $M_{n+1}$ obtained is as expected.
    \item[$\Rightarrow$)] Assume a transition $q_n,M_n\uberTo{x}
      q_{n+1},M_{n+1}$ is labeled with $x_{n+1}\notin Dom(M_n)$. By
      induction $x_{n+1}\in X_n$, and thus
      $(q_{n+1},X_{n+1})\in\delta' ( (q_n,X_n) , x_{n+1})$. \qed
    \end{trivlist} 
  \end{trivlist}
  Thus, for every run starting from an initial state and reaching a
  configuration $(q,X),M$ the couple $(dom(M),X)$ is a partition of
  $\mycal{X}$. Consequently each transition of $\mycal{A}'$ is labeled
  with a variable which is free in every run reaching its source
  state. Thus it suffices to prove that in $\mycal{A}'$, for every
  $n\ge 0$, there exists a path from an initial state to a final state
  of length $n$. We reduce this problem to the universality of the FA
  $A''$ on a unary alphabet $\set{a}$ obtained by replacing every
  transition $q_1\uberTo{x}q_2$ of $\mycal{A}'$ by the transition
  $q_1\uberTo{a}q_2$, where $a$ is an arbitrary letter in $\Sigma$.
  That is, we check whether $L(A'')=a^{\star}$.\qed 
\end{trivlist}
\qed



We cannot check
$L(\mycal{A}_1)\subseteq L(\mycal{A}_2)$ by intersecting
$L(\mycal{A}_1)$ with $\Sigma^* \setminus L(\mycal{A}_2)$ since the
latter is not necessarily a FVA language even when $\mycal{A}_2$ is a
FA.
However containment is decidable if one of the FVAs is a finite
automaton, since in this case the intersection of the languages is regular (Lemma~\ref{FA:inter:FVA:lemma} in the Appendix).

\begin{theorem}
  The containment problems between a FVA and a FA are decidable.
\end{theorem}


\section{Games for the simulation of communicating  FVAs}
\label{simulation:sec}
To deal with service composition problems we need first to extend FVAs
to the \emph{communicating FVAs}, or CFVA for short, where labels
(letters or variables) are prefixed by a communication symbol "!" or
"?".  Then we generalize the standard FA simulation relation to a FVA
simulation in order to formalize that a client can be satisfied by an
available service (when both are specified by a CFVA).  A client
transition labeled by $!x$, where $x$ is not bound, should be simulated by
a service transition which is labeled by $?y$, where $y$ is not bound as
well, since the service should handle all instances of $x$.  On the
other hand, a client transition labeled by $?x$, where $x$ is not bound,
can be simulated by a service transition labeled by any $!\alpha$.
Hence,  in order to define properly the simulation  we should take into account 
the refreshing of variables. 

\paragraph{Definition of CFVAs.}
Formally, a CFVA is defined exactly like a FVA but for the transition
function $\delta =Q\times \ccons \rightarrow 2^{Q}$, where for a set
$S$, $S^{!?}$ denotes the set $\set{!s,?s \,|\, s \in S}$.
To simplify the presentation from now we shall only consider 
CFVAs in which there is a unique initial state
and all the states are accepting. 
The definition of the simulation game for CFVAs follows.

\begin{definition}
\label{sim:game:def}
Let
$\mycal{A}_1=\model{\Sigma,\mathcal{X}_1,Q_1,q^{1}_0,\delta_1,F_1,\kappa_1}$
and
$\mycal{A}_2=\model{\Sigma,\mathcal{X}_2,Q_2,q^2_0,\delta_2,F_2,\kappa_2}$
be two CFVAs where $\mycal{X}_1 \cap \mycal{X}_2=\emptyset$.  
Let $\Pos$ be the set of positions reachable from
$p^{\star}=\AbelardM{(\emptyset,q^1_0),(\emptyset,q^2_0)}$ by the set
of moves $M=M_{A}^! \cup M_{A}^? \cup M_{E}^! \cup M_E^?$, where:
$$
\begin{array}{rl}
  M_A^?&=\big\{ \AbelardM{(\sigma_1,q_1), \varrho_2}
  \uberTo{} \EloiseM{({\sigma_1}_{{|D}},q'_1),  \varrho_2}{(\sigma_1,?\alpha)}\\
   &\hspace*{1cm}\vert\hspace*{0.5cm} q'_1 \in \delta_1(q_1,?\alpha)  \tand D=Dom(\sigma_1) \setminus \kappa_1^{-1}(q'_1) \big\}\\
  M_A^!&=\big\{ \AbelardM{(\sigma_1,q_1), \varrho_2} \uberTo{}  
  \EloiseM{((\sigma_1 \uplus \gamma)_{|D},q'_1),  \varrho_2}{(\gamma
    \uplus \sigma_1,!\alpha)} \\ 
  &\hspace*{1cm}\vert\hspace*{0.5cm}  q'_1 \in \delta_1(q_1,!\alpha)\\
  &\hspace*{1.5cm} \tand D=Dom(\sigma_1 \uplus \gamma) \setminus \kappa_1^{-1}(q'_1)\\  
  &\hspace*{1.5cm} \tand \gamma: \mycal{V}(\sigma_1(\alpha))\to \Sigma \big\} \\ 
  M_E^!&=\big\{\EloiseM{\varrho_1,(\sigma_2,q_2)}{(\sigma_1,!\alpha)}
  \uberTo{} \AbelardM{\varrho_1, ((\sigma_2 \uplus \sigma)_{|D},q'_2)}\\
  &\hspace*{1cm}\vert\hspace*{0.5cm}  q'_2 \in \delta_2(q_2,?\beta)\\
  &\hspace*{1.5cm} \tand D=Dom(\sigma_2 \uplus \sigma) \setminus \kappa_2^{-1}(q'_2)\\
  &\hspace*{1.5cm} \tand \sigma(\sigma_2(\beta))={\sigma_1(\alpha)}  \big\}\\
  M_E^?&=\big\{
  \EloiseM{(\sigma_1,q_1),(\sigma_2,q_2)}{(\sigma'_1,?\alpha)}
  \uberTo{} \AbelardM{((\sigma_1\uplus\sigma)_{|D_1},q_1),({(\sigma_2\uplus\gamma)}_{|D_2},q'_2)}  \\
  &\hspace*{1cm}\vert\hspace*{0.5cm}  q'_2 \in \delta_2(q_2,!\beta)  \\
  &\hspace*{1.5cm} \tand D_1=Dom(\sigma_1\uplus \sigma) \setminus \kappa_1^{-1}(q_1),\\
  &\hspace*{1.5cm} \tand D_2=Dom(\sigma_2 \uplus \gamma)\setminus \kappa_2^{-1}(q'_2) \\
  &\hspace*{1.5cm} \tand \sigma(\sigma'_1(\alpha))={\gamma(\sigma_2(\beta))}\\ 
  &\hspace*{1.5cm} \tand \gamma: \mycal{V}(\sigma_2(\beta))\to \Sigma\big\}\\  
\end{array}
$$
where the moves in $M_E^?\cup M_E^!$ are \textit{wrt} any possible
substitution $\sigma$.

We let $\Pos_E=\Pos\cap (\zeta_{\mycal{X}_1} \times Q_1) \times
(\zeta_{\mycal{X}_2} \times Q_2) \times (\zeta_{\mycal{X}_1} \times
\ccons)$ and $\Pos_A = \Pos \cap (\zeta_{\mycal{X}_1} \times Q_1)
\times (\zeta_{\mycal{X}_2} \times Q_2)$.  The \emph{simulation game} of
$\mycal{A}_1$ by $\mycal{A}_2$, denoted by
$\mycal{G}(\mycal{A}_1,\mycal{A}_2)$, is the two-players game $\langle
\Pos_{E},\Pos_{A},M, p^{\star} \rangle$. As usual, any infinite play is
winning for \Eloise, and any finite play is losing for the player who
cannot move.
\end{definition}

\begin{definition}
\label{sim:relation:def}
Let   $\mycal{A}_1=\model{\Sigma,\mathcal{X}_1,Q_1,q^{1}_0,\delta_1,F_1,\kappa_1}$
 and  $\mycal{A}_2=\model{\Sigma,\mathcal{X}_2,Q_2,q^2_0,\delta_2,F_2,\kappa_2}$ be two CFVAs.
There is a \gsim{simulation} of $\mycal{A}_1$ by $\mycal{A}_2$
 iff \Eloise  has a winning strategy in the game $\mycal{G}(\mycal{A}_1,\mycal{A}_2)$,
and we shall write  $\mycal{A}_1\mysim \mycal{A}_2$.
\end{definition}

\paragraph{Explanations of the rules of the game.} 
The simulation  game $\mycal{G}(\mycal{A}_1,\mycal{A}_2)$ 
 is played  between two players: \Abelard ($\forall$ or attacker) and 
  \Eloise ($\exists$  or defender). 
Its positions  are either of the  form $\AbelardM{(\sigma_1,q_1), (\sigma_2,q_2)}$ or 
$\EloiseM{(\sigma_1,q_1), (\sigma_2,q_2)}{(\sigma,\alpha)}$, where $\sigma_1,\sigma_2,\sigma$ 
are ground  substitutions, $q_1$ (resp. of $q_2$) is a state of $\mycal{A}_1$ (resp. $\mycal{A}_2$), 
  and    $\alpha$ is  a message in $\ccons$.
They correspond to  \Abelard positions (\texttt{A}) or  
\Eloise positions (\texttt{E}).
The moves $M_A^{?}$  state  that \Abelard chooses  a transition  $q_1\uberTo{?\alpha}q'_1$ 
 in $\mycal{A}_1$   and asks \Eloise to match it.
Consequently, all the variables that must be refreshed in the resulting state $q'_1$ are released.  
The moves $M_A^{!}$ are the same as $M_A^{?}$ apart that they deal with a 
 sending message $!\alpha$.
 In this case,  \Abelard must first instantiate the  variable in  $!\alpha$ (if any) with a letter  by a ground 
substitution $\gamma$,  then asks \Eloise to match the message  $\gamma(!\alpha)$. 
The moves $M^{!}_E$ state that   \Eloise 
 chooses  a transition $q_2\uberTo{?\beta} q'_2$ in $\mycal{A}_2$ 
to match the  message $\sigma_1(!\alpha)$. Indeed, she matches $\sigma_2(\beta)$
with $\sigma_1(\alpha)$ where $\sigma_2$ represents the   value of the variables in the state $q_2$.
The resulting  substitution $\sigma$ is stored in the resulting state $q'_2$, 
and all the variables that must be refreshed at $q'_2$ are released.  
The moves $M^{?}_E$ are like  $M^{!}_E$  except that \Eloise must first instantiate 
the possible variable of the sending message $\sigma_2(\beta)$ with a ground substitution $\gamma$.

Notice that for every \Eloise position $\EloiseM{(\sigma_1,q_1),(\sigma_2,q_2)}{(\sigma,\alpha)} \in \Pos_{E}$,
 the substitutions $\sigma_1$ and $\sigma$ coincide on $dom(\sigma_1)\cap dom(\sigma)$.
 Notice also that the simulation game might be infinite
 with possibly infinite branching since  $\Sigma$ is infinite. 

\emph{The \gsim{simulation} problem} for CFVAs is the following: 
 given two CFVAs  $\mycal{A}_1$ and  $\mycal{A}_2$,  is $\mycal{A}_1 \preceq \mycal{A}_2$? 
\begin{example}
Let $\mycal{A}$ and $\mycal{B}$ the CFVA depicted in the Figure \ref{sim:ex:vautomata}, 
where $\kappa(x)=\set{p_1}$ and $\kappa(y)=\set{p_0}$. 
One can show that  $\mycal{A} \mysim \mycal{B}$.
\begin{figure}[h*]
  \vspace{-22pt}
\begin{center}
\scalebox{0.8}{
\begin{tikzpicture}[shorten >=1pt,bend angle=45, node distance=2cm,on grid,auto, initial text=,
  bend angle=45] 
   \node[state,initial,accepting] (p_0)   {$p_0$}; 
   \node[state,accepting] (p_1) [right =of p_0] {$p_1$}; 
   \node[state,accepting](p_2) [right =of p_1] {$p_2$};
    \path[->] 
    (p_0) edge  node {$?x$} (p_1)
    (p_1) edge  node  {$?y$} (p_2)
    (p_2) edge  [bend left] node  {$?x$}  (p_1)
    (p_2) edge [bend right]  node [above] {$?z$} (p_0)
    -- (-0.7,0.6) node {$\mycal{A}$};
  
  \begin{scope}[xshift=8cm,yshift=0cm]
   \node[state,initial,accepting] (q_0)   {$q_0$}; 
   \node[state,accepting](q_1) [right=of q_0] {$q_1$};
    \path[->] 
    (q_0) edge  [loop above] node  {$!a$} ()
    (q_0) edge [bend left]  node  {$!b$} (q_1)
    (q_1) edge [bend left]  node  {$!c$} (q_0)
    -- (-0.7,0.6) node {$\mycal{B}$};
\end{scope}
\end{tikzpicture}}
\end{center}
  \vspace{-20pt}
\caption{CFVAs  $\mycal{A}$ and  $\mycal{B}$ with $\mycal{A}\preceq \mycal{B}$,  where $\kappa(x)=\set{p_1}$ and $\kappa(y)=\set{p_0}$.}
\label{sim:ex:vautomata}
  \vspace{-22pt}
\end{figure}

\end{example}

Ss\section{On the decidability of the \gsim{simulation} problem}
\label{termination:sec}
In this section  we  show that  the problem  of  \gsim{simulation} is decidable.
 The idea is that this problem can be reduced to a \gsim{simulation} problem 
over the same CFVAs
in which  the two players instantiate the variables  from a \emph{finite} set
 of letters, as proven in Proposition \ref{sigma:finite:prop}.

\begin{definition}
\label{sigma:finite:strategy:def}
Let
$\mycal{A}_1=\model{\Sigma,\mathcal{X}_1,Q_1,q^{1}_0,\delta_1,F_1,\kappa_1}$
and $\mycal{A}_2=\model{\Sigma,\mathcal{X}_2,Q_2,q^2_0,\delta_2,\\
  F_2,\kappa_2}$ be two CFVAs.  We define
$\overline{\mycal{G}}(\mycal{A}_1,\mycal{A}_2)$ to be the game
obtained by restricting the codomain of $\gamma$ to $C_0$ in the rules
of \Eloise $M_{E}^?$ and \Abelard $M_{\mycal{A}_1}^!$ in Def.
\ref{sim:game:def}, where $ C_0=\Sigma_{\mycal{A}_1} \cup
\Sigma_{\mycal{A}_2} \cup ({\mycal{X}_1 \times \mycal{X}_2}) \cup
({\mycal{X}_2 \times \mycal{X}_1})$.
\end{definition}
The following Lemma states an immediate   property of the game $\overline{\mycal{G}}$.
\begin{lemma}
\label{Gbar:finite:lemma}
Let $\mycal{A}_1,\mycal{A}_2$ be two CFVAs.
Then, the game $\overline{\mathcal{G}}(\mycal{A}_1,\mycal{A}_2)$ is
finite.
\end{lemma}
In order to prove  Proposition \ref{sigma:finite:prop} we need
to introduce the notion of coherence between substitutions and 
 between  game positions.

\begin{definition}
\label{coherence:subs:def}
 Let $C$ be a  finite subset of  $\Sigma$. 
The coherence relation $\synch_{C} \subseteq \zeta \times \zeta$ between 
   substitutions 
is defined by 
$\bar{\sigma} \synch_{_{C}} \sigma$  iff the three following conditions hold:
\begin{enumerate}
\item $dom(\bar{\sigma}) = dom(\sigma)$,
\item If $\bar{\sigma}(x) \in C$ then  $\bar{\sigma}(x)=\sigma(x) $, and 
     if  $\sigma(x) \in C$, then $\bar{\sigma}(x)=\sigma(x) $, for any variable $x \in dom(\sigma)$, and 
   \item for any variables $x,y \in  dom(\sigma)$,  $\bar{\sigma}(x)=\bar{\sigma}(y)$ iff $\sigma(x)=\sigma(y)$.
\end{enumerate} 
\end{definition}

\noindent The definition of the coherence between game positions, still denoted by $\synch_C$,
  follows.

\begin{definition}
\label{coherence:posi:def}
Let $C $ be a  finite subset of  $\Sigma$. \\
Let
$\mycal{A}_1=\model{\Sigma,\mathcal{X}_1,Q_1,q^{1}_0,\delta_1,F_1,\kappa_1}$
and
$\mycal{A}_2=\model{\Sigma,\mathcal{X}_2,Q_2,q^2_0,\delta_2,F_2,\kappa_2}$
be two CFVAs s.t. $\mycal{X}_1\cap \mycal{X}_2=\emptyset$.  
Let
$\Pos_E$ (resp. $\Pos_A$) be the set of \Eloise's (resp. \Abelard's)
positions in the game $\mycal{G}(\mycal{A}_1,\mycal{A}_2)$.  Then we
define the relation: $ \synch_{C} \; \subseteq \Pos_A\times \Pos_A
\,\cup \, \Pos_E\times \Pos_E $
by:  
    \begin{description} 
    \item[$\bullet$] For any $\sigma_i,\bar\sigma_i$ of proper domain
      included in $\mycal{X}_i$ ($i=1,2$) we have:\\
      $\big( \AbelardM{(\bar{\sigma}_1,q_1),(\bar{\sigma}_2,q_2)} \synch_{C} \AbelardM{(\sigma_1,q_1), 
   (\sigma_2,q_2)}\big)$ iff 
$   (\bar{\sigma}_1 \uplus \bar{\sigma}_2) \synch_{C}  (\sigma_1 \uplus \sigma_2)$.
\item[$\bullet$] For any $\sigma_i,\bar\sigma_i$ of proper domain
  included in $\mycal{X}_i$ ($i=1,2$), for any  substitutions
  $\sigma,\bar\sigma$ with proper domain included in $\mycal{X}_1$, we have: \\
  $((\bar{\sigma}_1 \cup \bar{\sigma}) \uplus \bar{\sigma}_2 )
  \synch_{C} (({\sigma}_1\cup {\sigma}) \uplus {\sigma}_2 )$ iff  \\$\big(
  \EloiseM{(\bar{\sigma}_1,q_1),(\bar{\sigma}_2,q_2)}{(\bar{\sigma},\alpha)}
  \synch_{C} \EloiseM{(\sigma_1,q_1), (\sigma_2,q_2)}{(\sigma,\alpha)}\big)$.
   \end{description} 
\end{definition}

\noindent Now we are ready to show that the games $\mycal{G}$ and $\overline{\mycal{G}}$ are equivalent 
in the following sense: 
\begin{proposition}
\label{sigma:finite:prop}
Let $\mycal{A}_1=\model{\Sigma,\mathcal{X}_1,Q_1,q^{1}_0,\delta_1,F_1,\kappa_1}$
 and  $\mycal{A}_2=\model{\Sigma,\mathcal{X}_2,Q_2,q^2_0,\delta_2,\\F_2,\kappa_2}$ be two CFVAs. 
 Then, \Eloise has a 
 winning strategy in $\mathcal{G}(\mycal{A}_1,\mycal{A}_2)$ iff she 
 has a  winning strategy in $\overline{\mathcal{G}}(\mycal{A}_1,\mycal{A}_2)$.
\end{proposition}
\begin{proof} 
Up to renaming of variables, we can assume that $\mycal{X}_1 \cap \mycal{X}_2=\emptyset$.
For the direction "$\Rightarrow$"  we show that out of a  winning strategy 
of \Eloise   in ${\mycal{G}}(\mycal{A}_1,\mycal{A}_2)$ we construct a  winning strategy 
for her in $\overline{\mycal{G}}(\mycal{A}_1,\mycal{A}_2)$.
  For this purpose, we   show that each move of \Abelard in  
$\overline{\mycal{G}}(\mycal{A}_1,\mycal{A}_2)$ can be  mapped to an   \Abelard move in $\mycal{G}(\mycal{A}_1,\mycal{A}_2)$, 
and that  \Eloise response in $\mycal{G}(\mycal{A}_1,\mycal{A}_2)$ can be actually  mapped  to 
an  \Eloise move in $\overline{\mycal{G}}(\mycal{A}_1,\mycal{A}_2)$.
This mapping defines a relation $\mycal{R}$ 
 between the positions of $\overline{\mycal{G}}(\mycal{A}_1,\mycal{A}_2)$
  and the positions of $\mycal{G}(\mycal{A}_1,\mycal{A}_2)$ as follows:  
$  \mycal{R} \subseteq   \Pos_{E}(\overline{\mycal{G}}(\mycal{A}_1,\mycal{A}_2)) \times \Pos_{E}(\mycal{G}(\mycal{A}_1,\mycal{A}_2)) \cup
   \Pos_{A}(\overline{\mycal{G}}(\mycal{A}_1,\mycal{A}_2)) \times \Pos_{A}(\mycal{G}(\mycal{A}_1,\mycal{A}_2)), 
 $
such that if $(\bar{\wp},\wp) \in \mycal{R}$, 
 and the move $\bar{\wp}\uberTo{}\bar{\wp}'$  in 
  $\overline{\mycal{G}}(\mycal{A}_1,\mycal{A}_2)$ is mapped to $\wp \uberTo{} \wp'$ in $\mycal{G}(\mycal{A}_1,\mycal{A}_2)$, 
  or  $\wp \uberTo{} \wp'$ in $\mycal{G}(\mycal{A}_1,\mycal{A}_2)$ 
is mapped to $\bar{\wp}\uberTo{}\bar{\wp}'$ in $\overline{\mycal{G}}(\mycal{A}_1,\mycal{A}_2)$, 
 then $(\bar{\wp}',\wp') \in \mycal{R}$. 
 Furthermore, we impose that the following invariant (Inv-$\synch$) holds:
  If $(\bar{\wp}, \wp) \in  \mycal{R}$  
 then  $\bar{\wp} \synch_{C} \wp$,
where  $C=\Sigma_{\mycal{A}_1}\cup \Sigma_{\mycal{A}_2}$.
We recall that the variables in $\overline{\mycal{G}}(\mycal{A}_1,\mycal{A}_2)$  are instantiated from the set of letters
 $C_0= \Sigma_{\mycal{A}_1} \cup \Sigma_{\mycal{A}_2} \cup ({\mycal{X}_1 \times  \mycal{X}_2}) 
  \cup  ({\mycal{X}_2 \times \mycal{X}_1})$.
The main part of the proof consists  in  finding the right way to relate the instantiation of 
 the variables of the sending messages in  $\overline{\mycal{G}}(\mycal{A}_1,\mycal{A}_2)$ 
and $\mycal{G}(\mycal{A}_1,\mycal{A}_2)$. More precisely, we distinguish three cases:
when \Abelard in $\overline{\mycal{G}}(\mycal{A}_1,\mycal{A}_2)$ instantiates a variable with a letter 
 in  $\Sigma_{\mycal{A}_1} \cup \Sigma_{\mycal{A}_2}$, then 
 \Abelard in  $\mycal{G}(\mycal{A}_1,\mycal{A}_2)$ must instantiate the same  variable with the same  letter.
When \Abelard in $\overline{\mycal{G}}(\mycal{A}_1,\mycal{A}_2)$ instantiates a variable with a \emph{fresh} letter that  belongs to  
$C_0\setminus(\Sigma_{\mycal{A}_1} \cup \Sigma_{\mycal{A}_2})$ --by fresh we mean  it does not appear in the current position of $\overline{{\mycal{G}}}(\mycal{A}_1,\mycal{A}_2)$--
 then 
 \Abelard in  $\mycal{G}(\mycal{A}_1,\mycal{A}_2)$ must instantiate the same  variable with a fresh letter in  $\Sigma$.
Finally,  when \Abelard in $\overline{\mycal{G}}(\mycal{A}_1,\mycal{A}_2)$
 instantiates a variable with a \emph{non fresh} letter, say $\bar{a}_0$, i.e. 
$\bar{a}_0$  appears  in the current position, then 
 \Abelard in  $\mycal{G}(\mycal{A}_1,\mycal{A}_2)$ must instantiate the same  variable with the  letter $a_0$ related to $\bar{a}_0$, 
 i.e. in a previous step the choice of $\bar{a}_0$ corresponds to the choice of $a_0$.
For the other direction, i.e. \Eloise instantiation of the variables in $\mycal{G}(\mycal{A}_1,\mycal{A}_2)$ from $\Sigma$
is related to \Eloise instantiation of the variables in $\overline{\mycal{G}}(\mycal{A}_1,\mycal{A}_2)$ from $C_0$ by 
following the same principle. Following this construction, we ensure that 
 the  invariant (Inv-$\synch$) is always maintained.

\noindent The proof of the direction ($\Leftarrow$) is similar to the one of 
 ($\Rightarrow$): we follow the  same instantiation principle and keep the 
  same  definition of the $\synch$-coherence.\qed
\end{proof}
It follows from  Lemma \ref{Gbar:finite:lemma} and Proposition \ref{sigma:finite:prop}: 
\begin{theorem}
\label{main:decidable:th}
The problem of \gsim{simulation} is decidable  for  CFVAs.
\end{theorem}
Given two CFVAs $\mycal{A}_1,\mycal{A}_2$, deciding whether $\mycal{A}_1 \preceq \mycal{A}_2$ 
simply amounts to construct the finite game $\overline{\mathcal{G}}(\mycal{A}_1,\mycal{A}_2)$ and 
compute a winning strategy for \Eloise. 





\section{Service  composition} 
\label{service:synthesis}
To carry on the \ttt{CART} example and real-world service
applications, we need to extend CFVAs and \gsim{simulation} so that
transitions labels can be of type $!t$ or $?t$, with $t$ an arbitrary
term over a first-order signature. This extended model (ECFVA) is
detailed in Appendix \ref{extensions:FVA:appendix}.
\gsim{simulation} problem remains decidable for the subclass of ECFVAs in
which the terms labeling the transitions are either constants or of
the form $f(\alpha_1,\ldots,\alpha_n)$ where $f$ is a functional
symbol and $\alpha_i$ is either a variable or a constant, as is the
case for the \ttt{CART} example.

\paragraph{Composition synthesis.} We consider the same composition
synthesis problem as in~\cite{MW08,BerardiCGP08} besides the modelling
of the client goal and each service as an ECFVA. We
adapt  the construction of the  asynchronous product $\otimes$ on FAs \cite{MW08} 
for ECFVAs  to obtain an ECFVA modelling the community of available services.  Finding a
simulation then amounts to constructing a winning strategy for \Eloise in the simulation game.
In the case of the \ttt{CART} example, one strategy can be computed 
in the  game $\mycal{G}(\ttt{CLIENT},\ttt{CART}\otimes \ttt{SEARCH})$,
and thus the client requests can be satisfied. 
%
Notice that this problem is EXPTIME-hard as a direct
consequence of \cite{MW08}, where this lower bound obtained for the
composition synthesis of deterministic finite automata is established.



\section{Conclusion}
\label{conclusion:sec}

In future works we plan to investigate the   complexity 
of the universality and  \gsim{simulation}  of CFVAs and 
to find   other  classes of ECFVAs  for which the \gsim{simulation}  can be decided.
It would be important to  consider security constraints  
that the composition of services must fulfill as in~\cite{SCC-WSCA08}. 
For this purpose, suitable model-checking techniques have to be devised for  FVAs.

\bibliographystyle{abbrv}
\bibliography{biblio}

   \newpage
   \appendix 
\section*{Appendices}
\label{annex}

\setcounter{theorem}{0}
\setcounter{proposition}{0}

\section{On the comparison with other models}
FVAs are incomparable with variable automata \cite{GKS10}. On the one hand the language 
$L=\set{a_1a_1a_2a_2\cdots a_n a_n, n\ge 0, a_i \in \Sigma}$ cannot be   
recognized by a variable automaton as shown in \cite{GKS10}. However, 
 it is  recognized by the FVA $\mycal{A}_1$  of Example \ref{ex:VFA}. 
On the other hand,  the language of all the words in 
which the last letter is different from all the other letters 
can be recognized by a variable automaton but not by a FVA,
since there is no way to express in FVAs that a variable  is distinct from  
other variables.  Besides,  the subclass of variable automata without free variables 
  coincides with the subclass of FVAs without fresh variables.

FVAs are weaker than   FMAs  \cite{tcs/KaminskiF94}. 
The language of  words  in which some 
letter appears exactly twice can be recognized by a FMA  \cite{tcs/KaminskiF94} 
but not by a FVA. 


\section{Appendix for Section \ref{properties:FVA:sec}}
\label{closure:sec:appendix}

Before establishing the proofs of the claims of Section \ref{properties:FVA:sec}, 
we first give the formal definition of configuration and run for $n$-BFVAs
since it is required thereafter. 

\subsection{Run and configuration for $n$-BFVAs}
\begin{definition}
\label{run:lang:def}
Let   $\mycal{A}=\model{\Sigma,\mathcal{X},Q,Q_0,\delta,F,\kappa}$ be an $n$-FVA.  
A \emph{configuration}  is  a pair  $(q,M)$ where $q \in Q$ and 
$M:\mycal{X} \pfun  \Sigma$  is a substitution.
 We define a transition relation  over the  configurations 
  as follows:  $(q_1,M_1)\uberTo{u} (q_2,M_2)$, where $u\in \Sigma$,
   iff there exists an $n$-label  $\bs{l}_n=(l_1,\ldots,l_n)  \in (\Sigma \cup \mycal{X})^n$, 
  such that  $q_2\in \delta(q_1,\bs{l}_n)$, and a substitution $\sigma: \mycal{X}\rTo \Sigma$ such that 
  $\sigma(M_1(l_i))=u$,  
  for all $i \in \set{1,\ldots n}$,
  so that $M_2= (M_1 \uplus \sigma)_{| D}$, where $D=Dom(M_1 \uplus \sigma) \setminus  \kappa^{-1}(q_2)$.
   A finite word $u=u_1u_2\ldots u_m \in {\Sigma}^*$ is \emph{recognized} by 
$\mycal{A}$ iff  there exists a run
$(q_0,M_{0}) \uberTo{u_1}(q_1,M_1)\uberTo{u_2} \ldots \uberTo{u_m}(q_m,M_m)$, such that 
$M_{0}=\emptyset$, $q_0\in Q_0$ and $q_m \in F$.
 The set of words recognized by $\mycal{A}$ is still denoted by
$L(\mycal{A})$.
\end{definition}    
\begin{figure}[h*]
\begin{center} 
\begin{tikzpicture}[shorten >=1pt,node distance=2cm, bend angle=60,
   on grid,auto, initial text=, >=stealth] 
  \begin{scope}[xshift=0cm, yshift=0cm,node distance=2cm]
   \node[state,initial,accepting] (r_0)   {$q_0$}; 
   \node[state] (r_1) [right =of r_0] {$q_1$}; 

    \path[->] 
    (r_0) [bend left] edge   node [above] {$(a,y)$} (r_1)
    (r_1)  edge  node  [below] {$(x,y)$} (r_0)
    -- (-2,0) node {$\mycal{A}$};
 \end{scope}
\end{tikzpicture}
\end{center}
\caption{A $2$-FVA.}
  \vspace{-10pt}
\end{figure}

\begin{example}
Let $\mycal{A}$ be the $2$-FVA depicted below where $\kappa(y)=\set{q_0,q_1}$ 
and $\kappa(x)=\emptyset$. It is clear that $L(\mycal{A})=\set{(az)^n \,|\, z  \in \Sigma, n\ge 1}$.
\end{example}

\subsection{Closure under basic operations}
 \label{Intersec:sec:appendix}

 The class of FVAs with $\eps$-transitions will be denoted by $\eps$-FVAs.
\begin{lemma}
\label{eps:closed:lemma}
For a $\eps$-FVA $\mycal{A}^{\eps}$  there exists a FVA $\mycal{A}$ (without $\eps$-transitions)
satisfying $L(\mycal{A})=L(\mycal{A}^{\eps})$.
\end{lemma}
\begin{proof}
The construction of a FVA  out of a $\eps$-FVA 
 is more subtle than the  construction known for FAs 
 since we need to take into account the refreshing of the variables.
We define an operator $\Theta$ that transforms  a $\eps$-FVA  
to an  equivalent $\eps$-FVA with strictly less $\eps$-transitions. Thus the desired 
    FVA without $\eps$-transitions is the least fixed-point of $\Theta$.
Intuitively, the operator $\Theta$ eliminates all the $\eps$-transitions 
which are preceded by a non $\eps$-transition.   

  Assume $\mycal{A}^{\eps}=\model{\Sigma,\mycal{X},Q^{\eps},Q_0^{\eps},F^{\eps},\delta^{\eps},\kappa^{\eps}}$.
Let $\Upsilon(q)$ be the set of states that are reachable from state $q$ by following an $\eps$-transition
 and let $\Upsilon(Q')=\set{\Upsilon(q)| q \in Q'}$, for  $Q' \subseteq Q^\eps$. 
  Let
  $\Theta(\mycal{A}^{\eps})=\model{\Sigma,\mycal{X},Q,Q_0,F,\delta,\kappa}$
  where:
  \begin{align*}
    Q  &= Q^\eps \cup (Q^\eps \times Q^\eps) \\
    & \pi_1:\mycal{P}(Q) \to \mycal{P}(Q^\eps)\\
    & \quad \quad Q' \mapsto \set{ p \,\vert\, (p,q)\in Q'} \\    
    & \pi_2: \mycal{P}(Q) \to \mycal{P}(Q^\eps)\\
    & \quad \quad Q' \mapsto \set{ q \,\vert\, (p,q)\in Q'} \\    
    Q_0 &=Q_0^\eps \cup \Upsilon(Q_0^\eps) \cup \pi_1^{-1}(Q_0^\eps) \\
    F &=F^\eps \cup \Upsilon^{-1}(F^\eps) \cup \pi_2^{-1}(F^\eps)\\
    \delta&= \set{p\uberTo{\alpha}q \in \delta^\eps \gvert \alpha \neq \eps } \cup  
    \set{q_1\uberTo{\alpha}(q_2,q_3) \gvert q_1\uberTo{\alpha}q_2\uberTo{\eps} q_3 \in \delta^\eps \gvert \alpha \neq \eps} \cup \\
        & \quad \;    \set{(q_1,q_2)\uberTo{\alpha}q_3 \gvert q_2\uberTo{\alpha} q_3 \in \delta^\eps }
       \cup \set{q_1 \uberTo{\eps} q_2 \gvert {\nexists} q_0\uberTo{\alpha} q_1 \tst \alpha \neq \eps}\\
     \kappa&=\kappa^{\eps} \cup (\pi_1^{-1} \circ \kappa^\eps) \cup (\pi_2^{-1} \circ \kappa^\eps)
  \end{align*}

In order to prove that $L(\Theta(\mycal{A}^{\eps}))=L(\mycal{A}^{\eps})$, it suffices to prove 
 the following three Claims, the first one is straightforward:

\begin{trivlist}
  \item \textbf{Claim 1.} \it Every accepting run in $\mycal{A}^{\eps}$
    that does not follow any $\eps$-transition is still an accepting run  
    in $\Theta(\mycal{A}^{\eps})$. Conversely,  every accepting run in $\Theta(\mycal{A}^{\eps})$
    that passes only through states in $Q^{\eps}$ is still an accepting run in $\mycal{A}^\eps$.
\end{trivlist}

  \begin{trivlist}
  \item \textbf{Claim 2.} \it There exists a run 
    \begin{align*}
      q_0,M_0   \uberTo{\alpha} q_1,M_1  \uberTo{\eps} q_{2},M_{2} 
    \end{align*}
  in $\mycal{A}^{\eps}$ with $\alpha \neq \eps$ 
    \emph{iff}  there exists a run
    \begin{align*}
      q_0,M_0 \uberTo{\alpha}  (q_1,q_2),M'_2
    \end{align*}
in $\Theta(\mycal{A}^{\eps})$ such that  $M_2=M'_2$.

  \end{trivlist}
  \begin{trivlist}
  \item \textbf{Proof of the Claim.}
    \begin{itemize}
    \item[$\Rightarrow$)] From the definition of $Q$ and $\delta$ it follows that 
      $(q_0,q_1) \in \delta(q_0,\alpha)$, and it remains to show that $M_2=M'_2$. 
      We only discuss the case when $\alpha$ is a letter in $\Sigma$, the case when it 
      is a variable can be handled similarly. 
      On the one hand, $M_2={M_1}_{|D_2}$ where $D_2=Dom(M_1) \setminus  (\kappa^{\eps})^{-1}(q_2)$, 
      and $M_1={M_2}_{|D_1}$ where $D_1=Dom(M_0) \setminus  \kappa^{-1}(q_1)$. Hence
      $M_2={M_0}_{|D}$ where  $D=Dom(M_0) \setminus  \big((\kappa^{\eps})^{-1}(q_1) \cup (\kappa^{\eps})^{-1}(q_2)\big)$.
On the other hand, we have $M'_2=M_0 \setminus D'$, where $D'=Dom(M_0) \setminus  \kappa^{-1}((q_1,q_2))$.
  It follows from the definition of $\kappa$, the refreshing function 
   of $\Theta(\mycal{A}^{\eps})$, that $\kappa^{-1}((q_1,q_2))=(\kappa^{\eps})^{-1}(q_1) \cup (\kappa^{\eps})^{-1}(q_2)$.
   Hence, $D=D'$ and $M_2=M'_2$.

    \item[$\Leftarrow$)] This direction is proved by following the same reasoning made in   the direction $(\Rightarrow)$
   on the      refreshing  function. 
 \end{itemize} 
This ends the proof of Claim 2. \qed

  \begin{trivlist}
  \item \textbf{Claim 3.} \it Let $q_1  \in Q^{\eps}$  and $(q_0,q_1) \in Q$. There exists a run 
    \begin{align*}
      q_1,M_1   \uberTo{\alpha} q_2,M_2 
    \end{align*}
  in $\mycal{A}^{\eps}$ 
    \emph{iff}  there exists a run
    \begin{align*}
      (q_0,q_1),M_1 \uberTo{\alpha}  q_2,M_2
    \end{align*}
in $\Theta(\mycal{A}^{\eps})$.
  \end{trivlist}
 \begin{trivlist}
  \item \textbf{Proof of the Claim.}
 By checking the transition function $\delta$. \qed
  \end{trivlist}
To accomplish the proof, it remains to notice that 
 if $q \in Q$ is such that $q \notin \pi_{1}^{-1}(Q^{\eps})$, then 
 the outgoing transitions  from $q$ in $\mycal{A}^{\eps}$  are exactly 
  the outgoing transitions from $q$ in $\Theta(\mycal{A}^{\eps})$.
\end{trivlist}
\qed
\end{proof}


 \begin{lemma}
  \label{2FVA:FVA:lem}
  $2$-FVAs and FVAs  are equivalent (i.e. recognize the same languages). 
 \end{lemma}
 
\begin{proof}
  First it is trivial that any language recognized by a FVA
  $\mycal{A}$ is also recognized by the 2-FVA $\mycal{A}'$, a copy
  of $\mycal{A}$ in which transitions are indexed by couples $(x,x)$
  instead of a variable or constant $x$.

  Now let $L$ be a language recognized by a 2-FVA $\mycal{A}$, and
  let us construct a FVA $\mycal{B}$ that recognizes $L$. It
  suffices to prove that for any word $\omega$ there is a run of
  $\mycal{A}$ that ends in a final state if, and only if, there is a
  run of $\mycal{B}$ that also ends in a final state. In order to
  construct $\mycal{B}$ we first construct from $\mycal{A}$ another
  2-FVA $\mycal{A}'$ that recognizes the same language, and such
  that the translation of $\mycal{A}'$ into a 1-FVA is trivial. In
  order to simplify notations, we assume in this proof that the
  assignment $M$ on the variables of an automaton is extended by the
  identity function on the set $\Sigma_A$ of letters occurring in
the 2-FVA.

  \paragraph{Definition of $\Psi$.} Let
  $\mycal{A}=\model{\Sigma,\mathcal{X},Q,Q_0,\delta,F,\kappa}$, and
  let $n_{\mycal{X}} = \vert \mycal{X} \vert$ and $n_{\Sigma} = \vert
  \Sigma_{\mycal{A}} \vert$, and assume $\Sigma_{\mycal{A}}=\set{a_1,\ldots, a_{n_\Sigma}}$.
  Let $\Psi \subseteq \set{1,\ldots,n_{\mycal{X}}+n_{\Sigma}}^{\Sigma_{\mycal{A}}
    \cup \mycal{X}}$ be the set of functions from $\Sigma_{\mycal{A}} \cup
  \mycal{X}$ to $\set{1,\ldots,n_{\mycal{X}}+n_{\Sigma}}$ such that
  for every $\psi\in\Psi$ we have $\psi(a_k) = k$. Furthermore, given
  $D\subseteq \mycal{X}$ and $\psi\in\Psi$, we let $\psi^{D}$ be the
  subset of $\Psi$ of functions equal to $\psi$ on $(\Sigma_{\mycal{A}} \cup
  \mycal{X} )\setminus D$. Finally, given a substitution  $M$ on
  $\Sigma_{\mycal{A}} \cup \mycal{X}$ we let $\Psi_{M}$ be the subset of $\Psi$ of
  functions $\psi$ such that, for all $x,y\in \Sigma_A\cup dom(M)$, we have $M(x) =
  M(y)$ \emph{iff} $\psi(x) = \psi(y)$.

  \paragraph{Construction of $\mycal{A}'$.} We let $\mycal{A}'$ be the
  2-FVA automaton
  $\model{\Sigma,\mathcal{X},Q',Q_0',\delta',F',\kappa'}$ where:
  $$
  \left\lbrace
    \begin{array}{ll}
      Q' =& Q \times \Psi  \\
      \pi:& \mycal{P}(Q') \to \mycal{P}(Q)\\
      &  Q'' \mapsto \set{ q \,\vert\, (q,\psi)\in Q''} \\
    \end{array}
  \right.
  $$
  and $Q_0'=\pi^{-1}(Q_0)$, $F'=\pi^{-1}(F)$, and $\kappa'=
  \pi^{-1}\circ \kappa$. The transition relation $\delta'$ is defined
  as follows for all $(q_0,\psi_0) \in Q'$ and $\alpha,\beta \in
  \Sigma_{\mycal{A}} \cup \mycal{X}$: 
  $$
    \delta ( (q_0,\psi_0) , (\alpha,\beta) ) =
    \lbrace ( q_1 , \psi_1) \,\vert\, 
    q_1 \in \delta ( q_0 , (\alpha,\beta) ) 
    \text{ and } \psi_0(\alpha)=\psi_0(\beta) 
    \text{ and } \psi_1 \in \psi_0^{\kappa^{-1}(q_1)} \rbrace 
  $$
  \begin{trivlist}
  \item \textbf{Claim.} \it There exists a run $q_0,M_0
    \uberTo{(\alpha_1,\beta_1)} q_1,M_1 \to \ldots
    \uberTo{(\alpha_n,\beta_n)} q_n,M_n$ in $\mycal{A}$
    \emph{iff} for all $\psi_n \in \Psi_{M_n}$ there exists a run
    $(q_0,\psi_0),M_0
    \uberTo{(\alpha_1,\beta_1)} (q_1,\psi_1),M_1 \to \ldots
    \uberTo{(\alpha_n,\beta_n)} (q_n,\psi_n),M_n$ in $\mycal{A}'$.
  \end{trivlist}

  \begin{trivlist}
  \item \textbf{Proof of the claim.}
    We prove the two implications by induction on $n$. The case $n=0$
    is trivial in both cases, so let us focus on the induction step in
    each direction.
    \begin{itemize}
    \item[$\Leftarrow$)] We note that since $\psi_{M_n}$ is never
      empty, it suffices to prove the existence of the run in
      $\mycal{A}$ for one run in $\mycal{A}'$. We leave to the reader
      this verification given the definition of the transition
      function.
    \item[$\Rightarrow$)] Assume that for every run of length $n$ in
      $\mycal{A}$ and for every possible $\psi_n$ there exists a run
      as prescribed in $\mycal{A}'$. Using the above notations, let us
      extend a run of length $n$ with a transition to $q_{n+1}\in
      \delta(q_n, (\alpha_{n+1},\beta_{n+1}) )$, and let $M_{n+1}$ be
      the assignment to variables in $q_{n+1}$.  It suffices to prove
      that for every function $\psi_{n+1}\in \Psi_{M_{n+1}}$ there
      exists a function $\psi_{n}\in \Psi_{M_{n}}$ such that
      $(q_{n+1},\psi_{n+1}) \in \delta ( (q_n,\psi_n),
      (\alpha_{n+1},\beta_{n+1}) )$. 

      First let us prove that the subset of functions $\psi_n$ such
      that there is a transition from $(q_n,\psi_n)$ with the pair
      $(\alpha_{n+1},\beta_{n+1})$ is not empty. This set contains all
      the functions $\psi_n$ such that:
      $$
      \left\lbrace
        \begin{array}{l}
          x,y \in \Sigma_A\cup dom(M_n), \psi_n(x) = \psi_n(y) \Leftrightarrow M_n(x) = M_n(y)\\
          \psi_n(\alpha_{n+1}) = \psi_n(\beta_{n+1})\\
        \end{array}
      \right.
      $$
      Since the transition is feasible on $q_n$ we note that if both
      $\alpha_{n+1}$ and $\beta_{n+1}$ are in $\Sigma_A\cup dom(M_n)$ we must have
      $ M_n(\alpha_{n+1}) = M_n(\beta_{n+1})$, and thus the second
      condition is satisfied. Otherwise, say if $\alpha_{n+1}$ is not
      in $\Sigma_A\cup dom(M_n)$, any value is possible for $\psi_n(\alpha_{n+1})$,
      including the value $\psi_n(\beta_{n+1})$. Thus, there exists 
      some states $(q_{n+1},\psi_{n+1}) \in \delta'( (q_n,\psi_n),
      (\alpha_{n+1} , \beta_{n+1} )$ for some $\psi_n$.

      Second, let us prove that for every $\psi_{n+1}$ such that for
      every $\psi_{n+1}\in\Psi_{M_{n+1}}$ there exists a $\psi_n$ as
      above such that $(q_{n+1},\psi_{n+1}) \in \delta'( (q_n,\psi_n),
      (\alpha_{n+1} , \beta_{n+1} )$. On the one hand, if a variable
      $x$ is refreshed and by definition of the transition relation on
      $\mycal{A}'$, if $(q_{n+1},\psi_{n+1})$ is reached then for
      every $l \in \set{1,\ldots,n_\Sigma+n_{\mycal{X}}}$ there exists
      $\psi'_{n+1}$ equal to $\psi_{n+1}$ but on $x$, where
      $\psi'_{n+1}(x)=l$.  On the other hand, if $x$ is not refreshed,
      then all the possible values of $\psi_{n+1}(x)$ are also all the
      possible values of $\psi_{n}(x)$ for the $\psi_n$ on which the
      transition is possible. This is easily proved by considering the
      three cases $x\in dom(M_{n+1}) \cap dom(M_n)$, $x\in
      dom(M_{n+1}) \cap \set{\alpha_{n+1},\beta_{n+1}}$, and $x \notin
      dom(M_{n+1})$ (and thus not in $dom(M_n)$) and proving that in
      each case the condition:
      $$
      \forall x,y\in \Sigma_A\cup dom(M), \psi(x) = \psi(y) \Leftrightarrow M(x)=M(y)
      $$
      holds for $\psi$ and $M_{n+1}$ if it holds for the same $\psi$
      and $M_n$ as long as $\psi(\alpha_{n+1}) = \psi(\beta_{n+1})$.
    \end{itemize}

    \paragraph{Construction of a 1-FVA from $\mycal{A}'$.} From
    $\mycal{A}'$ one constructs the 1-FVA:
    $$
    \mycal{B}=\model{\set{a_1,\ldots,a_{n_\Sigma}},\set{x_{n_\Sigma+1},\ldots,x_{n_\Sigma +
          n_{\mycal{X}}}},Q',Q_0',\delta'',F',\kappa''}
    $$
    where, with $c_i$ denoting either $a_i$ if $1\le i \le n_\Sigma$
    or $x_i$ if $n_\Sigma+1\le i\le n_\Sigma + n_{\mycal{X}}$:
    \begin{itemize}
    \item $q'\in{}\delta''(q,c_i)$ if, and only if, $q'\in
      \delta'(q,(\alpha,\beta))$ where $q=(q_0,\psi)$ and
      $\psi(\alpha)=\psi(\beta)=i$.
    \item $\kappa''(x_i)$ is the set of $( q_0,\psi )$ such that
      $\psi^{-1}(i) \subseteq \kappa^{-1}(q_0)$.
    \end{itemize}
  \end{trivlist}
\end{proof}

\begin{theorem}
\label{nFVA:FVA:Th}
For all $n\ge 1$, the $n$-FVAs and FVAs  are equivalent (i.e. they recognize the  same languages). 
\end{theorem}
\begin{proof}
We  prove by induction on $n \ge 1$ that  the $(n+1)$-FVAs and the  $n$-FVAs  are equivalent.
 The base case $n=1$ follows from Lemma \ref{2FVA:FVA:lem}.
For the induction case we  transform  a  $(n+1)$-FVA  $\mycal{A}$ to  an equivalent 
  $n$-FVA by contracting the first and the second component  of the $(n+1)$-labels 
 of $\mycal{A}$ as in 
the proof of Lemma \ref{2FVA:FVA:lem} and keeping the remaining $n-1$ components unchanged.
 \qed
\end{proof}

\begin{theorem}
\label{closure:Th}
FVAs are  closed under union, 
concatenation,  Kleene operator and intersection. 
\end{theorem}
\begin{proof}
Up to variable renaming  it is sufficient to consider the  union, intersection and 
concatenation  of two FVAs that do not share variables. 

We recall that the closure under union is straightforward since we just take the disjoint union of the two FVAs. 
 The closure under  Kleene operation and
concatenation    is a direct consequence of the fact that  FVAs with 
 $\varepsilon$-transitions and FVAs recognize the same language, Lemma \ref{eps:closed:lemma}. 

The closure under intersection for FVAs is an immediate consequence of  Theorem~\ref{nFVA:FVA:Th}, since
the intersection of two FVAs amounts to computing their  Cartesian product, which is  a $2$-FVA.  
 Formally, let  $\mathcal{A}_1=\model{\Sigma_1,\mycal{X}_1,Q_1,q_0^1,\delta_1,F_1,\kappa_1}$ 
and $\mathcal{A}_2=\model{\Sigma_2,\mycal{X}_2,Q_2,q_0^2,\delta_2,F_2,\kappa_2}$ be two  FVAs, 
 where  $\mycal{X}_1 \cap \mycal{X}_2 = \emptyset$. The $2$-FVA $\mycal{A}_1\times \mycal{A}_2$ is defined by:
\begin{align*}
  \mycal{A}_1\times \mycal{A}_2=\model{ \Sigma_1 \cup  \Sigma_2, \mycal{X}_1 \cup \mycal{X}_2, Q_1\times Q_2, q_0^1 \times q_0^2,\delta, F_1 \times F_2,\kappa},
\end{align*}
where $\delta$ and $\kappa$ are defined by:
 \[
\begin{cases}
(q'_1,q'_2) \in \delta((q_1,q_2),(\alpha_1,\alpha_2)) \quad \textrm{iff} \quad q'_1 \in \delta_1(q_1,\alpha_1)
 \textrm{ and }   q'_2 \in \delta_2(q_2,\alpha_2).  \\
     (q_1,q_2) \in \kappa(x)  \quad  \textrm{iff} \quad   q_1 \in \kappa_1(x) \tor  q_2 \in \kappa_2(x).
\end{cases}
 \]
The closure under intersection for FVAs follows from Lemma \ref{2FVA:FVA:lem} and the following Fact:
\begin{fact}
Let $\mathcal{A}_1$ and $\mathcal{A}_2$ be FVAs. 
Then, $L(\mycal{A}_1) \cap  L(\mycal{A}_2) =   L(\mycal{A}_1\times \mycal{A}_2)$.
\end{fact}
This ends the proof of Theorem \ref{closure:Th}.\qed
\end{proof}

\begin{lemma}
\label{vaut:complement:prop}
FVAs are not closed under complementation.
\end{lemma}
\begin{proof}
As a counter example we  consider the
language $L=\set{a}$, with $a \in \Sigma$.
  The complement of $L$ is the language $L_2 \uplus L_1$ where 
  $L_2$ consists of all  the words of length greater (or equal) than $2$ and 
 $L_1$ consists  of all  the words of length   $1$ in which the letter differs from $a$, i.e. 
$L_2= \set{a_1 a_2 \ldots a_n \,|\, a_i \in \Sigma, n\ge 2}$ and 
$L_1=\set{a_1 \,|\,  a_1  \in \Sigma \setminus \set{a}}$.  
The language $L_2$ can be recognized by a FVA. In order to show that $L_1 \uplus L_2$ 
    is not  FVA-recognizable,  it suffices to show that  $L_1$ is not  FVA-recognizable. 
 Towards a contradiction: assume that $L_1$ can be 
 recognized by a FVA $\mycal{B}$ without $\eps$-transitions. 
Hence $\mycal{B}$ must contain transitions of length  $1$ linking  an initial state to an accepting state. 
On the one hand, each transition of $\mycal{B}$ can not be labeled by a variable, otherwise $\mycal{B}$
 could accept words not in $L_2$. On the other hand,  
 all the transitions of $\mycal{B}$ must be labeled by letters  in $\Sigma \setminus \set{a}$, 
 but this is impossible since $\Sigma$ is infinite.  \qed
\end{proof}

\subsection{Nonemptiness and membership}
\label{Emptiness:membership:sec:annex}
\setcounter{theorem}{2}
\begin{theorem}
\label{Emptiness:membership:th}
For FVAs,  Nonemptiness is NL-complete and Membership is  NP-complete.
\end{theorem}
\begin{proof}
For Nonemptyness, let  $\mycal{A}$ be  a FVA and let  $\mycal{F}(\mycal{A})$ be FA obtained
from $\mycal{A}$ by considering all the variables  as letters. 
Notice  that  $\mycal{F}(\mycal{A})$ is nonempty iff $\mycal{A}$ is nonempty. 
The complexity follows from the fact that FA nonemptiness is NL-complete.\qed
For Membership, consider  a FVA  $\mycal{A}$  and  a word $w$. 
 For the upper bound, a non deterministic polynomial algorithm 
guesses a   path  in  $\mycal{A}$ of length $|w|$ such that the final state is accepting, 
 then checks wether  the corresponding  run on $w$ is possible.
 The lower bound is shown by a reduction from the Hamiltonian cycle problem  for 
 digraphs as in the extended version of \cite{GKS10}.
\qed
\end{proof}

\subsection{Containment}
 \label{contain:sec:appendix}

\begin{lemma}
\label{FA:inter:FVA:lemma}
Let $\mycal{A}$ be a FVA and $F$ be a FA. Then, $L(\mycal{A})\cap L(F)$ is regular. 
If $L(\mycal{A})= L(F)$ then all the paths  of $\mycal{A}$ linking an initial state to a final state  are labeled  
with letters.
\end{lemma}
\begin{proof}
The first claim  follows from   the proof of Theorem \ref{closure:Th}:
the construction of $\mycal{A}\cap F$ 
yields a FVA in which  all the  transitions 
are labeled with letters. 

For the second claim, 
assume that the regular language $L(F)$ is over a finite alphabet $\Sigma_f$. 
Towards a contradiction: 
 Let $q_1\uberTo{a_1}\ldots \uberTo{a_m}q_m\uberTo{x}\ldots q_k$
 be  a path in $\mycal{A}$  such that $q_1$ (resp. $q_k$) is an initial (resp. final) state, 
$x$ is a variable, and for every  $i \le m$, 
$a_i$ is a letter. 
Indeed, this path recognizes  a  word  $\bs{w}=w_1\ldots w_k$ that does not 
belong to $L(F)$, e.g. by choosing $w_{m+1} \notin \Sigma_f$. This is a contradiction. 
\qed
\end{proof}

\begin{theorem}
  The containment problems between a FVA and a FA are decidable.
\end{theorem}
\begin{proof}
Let  $\mycal{A}$ be  FVA and $F$ be a FA.

For the inclusion $L(F) \subseteq  L(\mycal{A})  $, we check whether 
$ L(F) \cap  L(\mycal{A})=L(F)$. From Lemma \ref{FA:inter:FVA:lemma} it follows that 
the language $ L(F) \cap  L(\mycal{A})$  is regular and the  FA recognizing  it can be constructed.
Hence,  the inclusion above  amounts to checking the inclusion  of two FAs, which is 
decidable. 

 For the inclusion $L(\mycal{A}) \subseteq  L(F)  $, we check whether 
$L(\mycal{A}) \cap L(F)=L(\mycal{A})$. 
On the one hand, it follows from Lemma \ref{FA:inter:FVA:lemma} that 
 $L(\mycal{A})\cap L(F)$ is regular. 
On the other hand, it follows from Lemma \ref{FA:inter:FVA:lemma}  that 
all the (accessible) transitions of $\mycal{A}$ must be labeled with letters, since $L(\mycal{A})$ is regular.
Hence, the  inclusion  above amounts to checking the inclusion of two FAs.
\qed
\end{proof}


\section{Appendix for  Section \ref{termination:sec}}
\label{annex:decid:gsim:sec}

The claims in the following remark are not hard to prove.
\begin{remark}
\label{synch:req}
Let $C\subseteq \Sigma$ be a finite set of  letters,  $\bar{\sigma}$ and $\sigma$  two substitutions, 
 $x$ a variable,    and $a$ a letter in $C$.
The following hold.  
If  $\bar{\sigma} \synch_{C} \sigma$ then $|codom(\bar{\sigma})|= |codom(\sigma)|$ and 
 $\bar{\sigma}_{|D} \synch_{C} \sigma_{|D}$,  where  $D\subseteq Dom(\sigma)$. 
Consequently, if $(\bar{\sigma}_1\uplus\bar{\sigma}_2) \synch  ({\sigma}_1\uplus {\sigma}_2)$ with 
 $dom(\bar{\sigma}_i)=dom(\sigma_i)$, then $\bar{\sigma}_i \synch \sigma_i$, for $i=1,2$. 
\end{remark}

\begin{proposition}
\label{sigma:finite:prop}
Let $\mycal{A}_1=\model{\Sigma_0,\mathcal{X}_1,Q_1,q^{1}_0,\delta_1,F_1,\kappa_1}$
 and  $\mycal{A}_2=\model{\Sigma_0,\mathcal{X}_2,Q_2,q^2_0,\delta_2,F_2,\kappa_2}$ be two FVAs. 
 Then \Eloise has a 
 winning strategy in $\mathcal{G}(\mycal{A}_1,\mycal{A}_2)$ iff she 
 has a  winning strategy in $\overline{\mathcal{G}}(\mycal{A}_1,\mycal{A}_2)$.
\end{proposition}
\begin{proof}
Up to variables renaming, we can assume  that $\mycal{X}_1 \cap \mycal{X}_2=\emptyset$.
For the direction "$\Rightarrow$"  we show that out of a  winning strategy 
of \Eloise   in ${\mycal{G}}(\mycal{A}_1,\mycal{A}_2)$ we construct a  winning strategy 
for her in $\overline{\mycal{G}}(\mycal{A}_1,\mycal{A}_2)$.
  For this purpose, we shall  show that each move of \Abelard in  
$\overline{\mycal{G}}(\mycal{A}_1,\mycal{A}_2)$ can be  mapped to an   \Abelard move in $\mycal{G}(\mycal{A}_1,\mycal{A}_2)$, 
and   \Eloise response in $\mycal{G}(\mycal{A}_1,\mycal{A}_2)$ can be actually  mapped  to 
an  \Eloise move in $\overline{\mycal{G}}(\mycal{A}_1,\mycal{A}_2)$.
This mapping defines a relation $\mycal{R}$ \footnote{More precisely, if $(\bar{\wp},\wp) \in \mycal{R}$, 
 and the move $\bar{\wp}\uberTo{\overline{\mycal{G}}}\bar{\wp}'$ is mapped to $\wp \uberTo{\mycal{G}} \wp'$, 
  or  $\wp \uberTo{\mycal{G}} \wp'$ is mapped to $\bar{\wp}\uberTo{\overline{\mycal{G}}}\bar{\wp}'$, 
 then $(\bar{\wp}',\wp') \in \mycal{R}$}  
 between the positions of $\overline{\mycal{G}}(\mycal{A}_1,\mycal{A}_2)$
  and the positions of $\mycal{G}(\mycal{A}_1,\mycal{A}_2)$ as follows:  
 \begin{align*}
  \mycal{R} \subseteq \; & \Pos_{E}(\overline{\mycal{G}}(\mycal{A}_1,\mycal{A}_2)) \times \Pos_{E}(\mycal{G}(\mycal{A}_1,\mycal{A}_2)) \;\; \cup  \\
    &  \Pos_{A}(\overline{\mycal{G}}(\mycal{A}_1,\mycal{A}_2)) \times \Pos_{A}(\mycal{G}(\mycal{A}_1,\mycal{A}_2)) 
 \end{align*}
 Furthermore, we impose that the following invariant holds:
\begin{align}
\label{invariant-sim}
\tag{Inv-$\synch$}
\textrm{If }  (\bar{\wp}, \wp) \in  \mycal{R}  
  \textrm{ then } \bar{\wp} \synch_{C} \wp,
\end{align}
where $C=\Sigma_{\mycal{A}_1}\cup \Sigma_{\mycal{A}_2}$.
In this proof,  we shall simply write  ``$\synch$" instead of ``$\synch_{C}$". 
We recall that the variables in $\overline{\mycal{G}}(\mycal{A}_1,\mycal{A}_2)$  are instantiated from the set of letters 
 $C_0= \Sigma_{\mycal{A}_1} \cup \Sigma_{\mycal{A}_2} \cup ({\mycal{X}_1 \times  \mycal{X}_2}) 
  \cup  ({\mycal{X}_2 \times \mycal{X}_1})$.
  The proof is by induction on $n$,  the number  of the  moves  
 made in $\overline{\mycal{G}}(\mycal{A}_1,\mycal{A}_2)$  plus the number of moves made in  ${\mycal{G}}(\mycal{A}_1,\mycal{A}_2)$.
The base case, i.e. when $n=0$,   trivially holds  since the starting position of 
$\overline{\mycal{G}}(\mycal{A}_1,\mycal{A}_2)$   and of $\mycal{G}(\mycal{A}_1,\mycal{A}_2)$ is 
  $\AbelardM{q_0^1,q_0^2}$.
 
For the induction case let $(\bar{\wp}_n,\wp_n) \in \mycal{R}$.  
We consider two possibilities: when  $\bar{\wp}_n$ and 
 $\wp_n$  are both \Abelard positions  and when they are both \Eloise positions. 
 Consider  the first  possibility and an  \Abelard move $\bar{m}=\bar{\wp}_n \uberTo{} \bar{\wp}_{n+1}$ in 
  $\overline{\mycal{G}}(\mycal{A}_1,\mycal{A}_2)$. 
 We distinguish two cases depending on  $\bar{m}$.

\begin{proofbycases}
\begin{caseinproof}
If  $\bar{m}\in M_{A}^?$, then $\bar{m}$ is of the form:
     \begin{align*} 
     \bar{m}= \AbelardM{(\bar{\sigma}_1,q_1), (\bar{\sigma}_2,q_2)} &\uberTo{}  
        \EloiseM{({\bar{\sigma}_1{}}_{|D},q'_1),  (\bar{\sigma}_2,q_2)}{(\bar{\sigma}_1,?\alpha)} \\
       &  \textrm{ where }  q'_1 \in \delta_1(q_1,?\alpha) \tand D=Dom(\bar{\sigma}_1) \setminus \kappa_1^{-1}(q'_1)
      \end{align*}
From  the induction hypothesis  we have  
  $\bar{\wp}_n \synch \wp_n$,  hence   $\wp_n=\AbelardM{(\sigma_1,q_1),(\sigma_2,q_2)}$ 
such that  $(\bar{\sigma}_1 \uplus \bar{\sigma}_2) \synch (\sigma_1\uplus \sigma_2)$.
 Thus \Abelard   move in $\mycal{G}(\mycal{A}_1,\mycal{A}_2)$ is 
\begin{align*} 
     \AbelardM{({\sigma}_1,q_1), (\sigma_2,q_2)} &\uberTo{}  
        \EloiseM{({{\sigma}_1}_{|D},q'_1),  (\sigma_2,q_2)}{(\sigma_1,?\alpha)} 
      \end{align*}
 and the invariant (\ref{invariant-sim}) is  maintained.
\end{caseinproof}

\begin{caseinproof}
If $\bar{m} \in M_{A}^!$, then $\bar{m}$ is of the form: 
\begin{align*}
\AbelardM{(\bar{\sigma}_1,q_1), (\bar{\sigma}_2,q_2)} &\uberTo{}  
        \EloiseM{((\bar{\sigma}_1 \uplus \bar{\gamma})_ {|D},q'_1), (\bar{\sigma}_2,q_2)}{(\bar{\gamma} \uplus \bar{\sigma}_1,!\alpha)}  \\
& \textrm{ where }  q'_1 \in \delta_1(q_1,!\alpha), D=Dom(\bar{\sigma}_1 \uplus \bar{\gamma}) \setminus \kappa_1^{-1}(q'_1)  \\
        &\tand \bar{\gamma}:\mycal{V}(\bar{\sigma}_1(\alpha))\to C_0 
\end{align*}
The only relevant situation is when $\bar{\sigma}_1(\alpha)$ is a variable, say $x_1\in \mycal{X}_1$. 
 The situation   when it  is a letter  is similar to the previous case since $\bar{\gamma}=\emptyset$. 
From the induction  hypothesis we have that $\bar{\wp}_n\synch \wp_n$, and hence 
$\wp_n=\AbelardM{({\sigma}_1,q_1), (\sigma_2,q_2)}$ such that $(\bar{\sigma}_1 \uplus \bar{\sigma}_2) \synch (\sigma_1\uplus \sigma_2)$. 
Therefore  the corresponding \Abelard move in   
$\mycal{G}(\mycal{A}_1,\mycal{A}_2)$ is 
\begin{align*}
\AbelardM{({\sigma}_1,q_1), (\sigma_2,q_2)} &\uberTo{}  
        \EloiseM{(({\sigma}_1 \uplus \gamma)_{|D},q'_1), (\sigma_2,q_2)}{(\gamma \uplus \sigma_1,!\alpha)} 
 \end{align*}
where   $\gamma: \mycal{V}(\sigma_1(\alpha)) \rightarrow  \Sigma$ is a (ground) substitution that will be defined next.
Since $\bar{\sigma}_1 \synch \sigma_1$, and $\bar{\sigma}_1(\alpha)$ is the  variable $x_1$, 
 then  it  follows  that $\bar{\sigma}_1(\alpha)=\sigma_1(\alpha)=\alpha=x_1$.  \Abelard choice of $\gamma$
depends on the nature of $\bar{\gamma}(x_1)$.
\begin{itemize}
\item  If $\bar{\gamma}(x_1)\in \Sigma_{\mycal{A}_1} \cup \Sigma_{\mycal{A}_2}$ then in this case
   we let $\gamma:=\bar{\gamma}$,  and hence  the invariant (\ref{invariant-sim}) is   maintained, i.e. 
 $(\bar{\sigma}_1\uplus \bar{\sigma}_2 \uplus  \bar{\gamma}) \synch ({\sigma}_1 \uplus {\sigma}_2 \uplus  \gamma)$.
\item  If $\bar{\gamma}(x_1)$ appears in the current position, i.e. 
   \begin{align*}
   \bar{\gamma}(x_1)  \in  (codom(\bar{\sigma}_1 \uplus  \bar{\sigma}_2)) \setminus (\Sigma_{\mycal{A}_1} \cup \Sigma_{\mycal{A}_2}),
   \end{align*}
 then  there is a variable $y \in dom(\bar{\sigma}_{1} \uplus \bar{\sigma}_2)$ such that
 $\big(y \mapsto \bar{\gamma}(x_1)\big) \in \bar{\sigma}_{1} \uplus \bar{\sigma}_2$. 
Since $(\bar{\sigma}_{1} \uplus \bar{\sigma}_2)  \synch ({\sigma}_{1} \uplus {\sigma}_2)$,
then it follows that there is a  letter $y_0\in \Sigma_0 $ such that  $\big(y \mapsto y_0 \big) \in {\sigma}_1 \uplus {\sigma}_2$.
Thus we let $\gamma:=\set{x_1 \mapsto y_0}$ and the invariant (\ref{invariant-sim}) is maintained, 
i.e.  $(\bar{\sigma}_1\uplus \bar{\sigma}_2 \uplus  \bar{\gamma}) \synch ({\sigma}_1 \uplus {\sigma}_2 \uplus  \gamma)$.  
\item  Otherwise, i.e. $\bar{\gamma}(x_1)$ is a new letter that does not appear in the current position,
      then we take $\gamma(x_1)$ as a new letter from $\Sigma_0$, and hence   the invariant (\ref{invariant-sim}) is maintained.  
\end{itemize}
\end{caseinproof}
\end{proofbycases}

Secondly, we  consider  the possibility  when 
 both $\bar{\wp}_n$ and  $\wp_n$  are  \Eloise positions.
We consider an  \Eloise move $m={\wp}_n \uberTo{} {\wp}_{n+1}$ in 
  ${\mycal{G}}(\mycal{A}_1,\mycal{A}_2)$, and 
 we describe  the corresponding  \Eloise move in   $\overline{\mycal{G}}(\mycal{A}_1,\mycal{A}_2)$. 
 We distinguish two cases depending on  $m$.

\begin{proofbycases}

\begin{caseinproof}
If $m \in M_{E}^{!}$, then  $m$ is of the form: 
\begin{align*}
 \EloiseM{ (\sigma_1,q_1), (\sigma_2,q_2)}{(\sigma_3,!\alpha)} &\uberTo{}  
        \AbelardM{(\sigma_1,q_1), ((\sigma_2 \uplus \sigma)_{|D},q'_2) }  \big\}\\
       &  \textrm{ where }  q'_2 \in \delta_2(q_2,?\beta), \\
      & D=Dom(\sigma_2\uplus \sigma) \setminus \kappa_2^{-1}(q'_2),  \tand  \\
      & \sigma({\sigma_2(\beta)})={\sigma_3(\alpha)}, \text{ for a substitution } \sigma
\end{align*}
Recall that $\sigma_3(\alpha)$ is a letter. 
From the induction hypothesis we have that $\bar{\wp}_{n} \synch \wp_{n}$, therefore
$\bar{\wp}_n= \EloiseM{(\bar{\sigma}_1,q_1), (\bar{\sigma}_2,q_2)}{(\bar{\sigma}_3,!\alpha)}$
such that $((\bar{\sigma}_1 \cup \bar{\sigma}_3) \uplus \bar{\sigma}_2 ) \synch (({\sigma}_1\cup {\sigma}_3) \uplus {\sigma}_2)$.
The corresponding move $\bar{m}$ in $\overline{\mycal{G}}(\mycal{A}_1,\mycal{A}_2)$
 is: 
\begin{align*}
 \EloiseM{ (\bar{\sigma}_1,q_1), (\bar{\sigma}_2,q_2)}{(\bar{\sigma}_3,!\alpha)} &\uberTo{}  
        \AbelardM{(\bar{\sigma}_1,q_1), ((\bar{\sigma}_2 \uplus \bar{\sigma})_{|D},q'_2)}, \\
\end{align*}
where $\overline{\sigma}$ is a (possibly trivial) 
substitution such that $\bar{\sigma}({\bar{\sigma}_2(\beta)})={\bar{\sigma}_3(\alpha)}$. 
But we    show that such a substitution exists 
 and that the invariant (\ref{invariant-sim}) is maintained. 
 Notice that $\bar{\sigma}_2(\beta)$ is a variable iff $\sigma_2(\beta)$ is a variable, and if so then  
  $\bar{\sigma}_2(\beta) = \sigma_2(\beta)$, since 
   $\bar{\sigma}_2 \synch \sigma_2$. Hence, 
   we shall show that the invariant is maintained  only  when $\sigma_2(\beta)$ and  $\bar{\sigma}_2(\beta)$ 
 are variables.  We distinguish 
 two  cases according to the nature of $\sigma_2(\beta)$:
 \begin{itemize}  
   \item If $\sigma_2(\beta)$ is a variable, say $x_2 \in\mycal{X}_2$, (i.e. $x_2 \notin dom(\sigma_2)$),
    then $\bar{\sigma}_2(\beta)=\sigma_2(\beta)=\beta=x_2$.
    We must show   that 
$(\bar{\sigma}_1 \uplus \set{ x_2 \mapsto \bar{\sigma}_3(\alpha)} \uplus \bar{\sigma}_2)
 \synch ({\sigma}_1 \uplus \set{ x_2 \mapsto {\sigma}_3(\alpha)} \uplus {\sigma}_2)$. Since 
we already know that  $(\bar{\sigma}_1 \cup \bar{\sigma}_3) \uplus \bar{\sigma}_2
 \synch ({\sigma}_1 \cup  {\sigma}_3) \uplus {\sigma}_2)$ then the claim  follows 
from the following fact:
\begin{fact}
Let  $\bar{\sigma}$ and $\sigma$ be two substitutions.
If $\bar{\sigma} \synch \sigma$,  and $x\in dom(\sigma)$ and $z\notin dom(\sigma)$, then
$\bar{\sigma}[z:=x] \synch \sigma[z:=x]$, where $\sigma[z:=x]$ stands for the replacement of
$x$ by $z$ in $\sigma$.
\end{fact}
   \item If $\sigma_2(\beta)$ is a letter,  then  $\sigma_2(\beta) = \sigma_3(\alpha)$. 
     We distinguish two cases depending on $\sigma_3(\alpha)$: 
     \begin{itemize}
     \item If $\sigma_3(\alpha) \in \Sigma_{\mycal{A}_1}\cup \Sigma_{\mycal{A}_2}$ 
    (and so $\sigma_2(\beta)$), then on the one hand,
     $\bar{\sigma}_3(\alpha)=\sigma_3(\alpha)$, 
       since $\bar{\sigma}_3 \synch \sigma_3$, and on the other hand,
       $\bar{\sigma}_2(\beta)=\sigma_2(\beta)$ since $\bar{\sigma}_2 \synch \sigma_2$.
       Therefore $\bar{\sigma}_3(\alpha)=\bar{\sigma}_2(\beta)$, and we are done. 
        \item  If $\sigma_3(\alpha) \in \Sigma \setminus(\Sigma_{\mycal{A}_1}\cup \Sigma_{\mycal{A}_2})$,  
   then $\alpha$ must be a  variable, say
           $x_1 \in \mycal{X}_1$. In this case 
           $\beta$ is also  a variable, say $x_2 \in \mycal{X}_2$,  since $\sigma_2(\beta)=\sigma_3(\alpha)$. 
   Notice that, on the one hand,  $\set{x_1 \mapsto \sigma_3(\alpha), x_2  \mapsto \sigma_3(\alpha)}$ appears in the 
   position $\wp_{n}$, i.e.    $\set{x_1 \mapsto \sigma_3(\alpha), x_2  \mapsto \sigma_3(\alpha)} \subset \sigma_1\cup \sigma_2\cup \sigma_3$.
   On the other hand, $\set{x_1 \mapsto \bar{\sigma}_3(\alpha), x_2  \mapsto \bar{\sigma}_2(\beta)}$ also appears in $\bar{\wp}_n$, i.e. \\
$\set{x_1 \mapsto \bar{\sigma}_2(\alpha), x_2  \mapsto \bar{\sigma}_3(\beta)} \subset \bar{\sigma}_1\cup \bar{\sigma}_2\cup \bar{\sigma}_3$.
     Therefore $\bar{\sigma}_2(\alpha)=\bar{\sigma}_3(\beta)$,       since $(\bar{\sigma}_1\cup \bar{\sigma}_2\cup \bar{\sigma}_3) \synch ({\sigma}_1\cup {\sigma}_2\cup {\sigma}_3)$.
     \end{itemize}     
 \end{itemize}
\end{caseinproof}

\begin{caseinproof}
If $m \in M^?_{E}$, then in this case this move is of the form
\begin{align*}
     \EloiseM{ (\sigma_1,q_1), (\sigma_2,q_2)}{(\sigma_3,?\alpha)} &\uberTo{}  
        \AbelardM{((\sigma_1 \uplus \sigma)_{|D_1},q_1), ((\sigma_2 \uplus \gamma)_{|D_2},q'_2) } \\
        & \textrm{ where }  q'_2 \in \delta_2(q_2,!\beta),  \\
       &  D_1= Dom(\sigma_1 \uplus  \sigma) \setminus \kappa_1^{-1}(q_1), \\
      & D_2 = Dom(\sigma_2 \uplus \gamma) \setminus \kappa_2^{-1}(q'_2), \\
& \sigma({\sigma_3(\alpha)})=\gamma(\sigma_2(\beta)), \tand \\
& \gamma: \mycal{V}(\sigma_2(\beta)) \rightarrow \Sigma.
\end{align*}
From the induction hypothesis we have that $\bar{\wp}_{n} \synch \wp_{n}$, therefore\\
 $\bar{\wp}_n=\EloiseM{(\bar{\sigma}_1,q_1), (\bar{\sigma}_2,q_2)}{(\bar{\sigma}_3,?\alpha)}$
such that $(\bar{\sigma}_1\cup \bar{\sigma}_3) \uplus \bar{\sigma}_2 \synch ({\sigma}_1\cup {\sigma}_3) \uplus {\sigma}_2$.
The corresponding  \Eloise move  in  $\overline{\mycal{G}}(\mycal{A}_1,\mycal{A}_2)$  is: 
\begin{align*}
     \EloiseM{(\bar{\sigma}_1,q_1), (\bar{\sigma}_2,q_2)}{(\bar{\sigma}_3,?\alpha)} &\uberTo{}  
        \AbelardM{ ((\bar{\sigma}_1 \uplus  \bar{\sigma})_{|D_1},q_1), ((\bar{\sigma}_2 \uplus \bar{\gamma})_{|D_2},q'_2) } \\
        & \textrm{ where }   \bar{\sigma}({\bar{\sigma}_3(\alpha)})={\bar{\gamma}(\bar{\sigma}_2(\beta))}
\end{align*}
   and the (ground) substitution $\bar{\gamma}: \mycal{V}(\bar{\sigma}_3(\alpha)) \rightarrow C_0$  
   by \Eloise will be defined next, 
provided that the invariant (\ref{invariant-sim}) is 
  maintained. Notice that maintaining this invariant does make sense only when $\sigma_3(\alpha)$ or $\sigma_2(\beta)$ is 
 a variable. 
The choice of $\overline{\gamma}$ depends on $\sigma_3(\alpha)$.
\begin{itemize}
\item If $\sigma_3(\alpha) \in \Sigma_{\mycal{A}_1} \cup \Sigma_{\mycal{A}_2}$, then this case is straightforward.  
\item If $\sigma_3(\alpha) \in \Sigma \setminus( \Sigma_{\mycal{A}_1} \cup \Sigma_{\mycal{A}_2})$, then
   $\alpha$ must be a variable,  say  $y_1 \in \mycal{X}_1$.
 We distinguish two cases depending on $\sigma_2(\beta)$. 
  \begin{itemize}
  \item If $\sigma_2(\beta)$ is a letter then in this case   
    $\sigma_2(\beta)=\sigma_3(\alpha)$, and  hence $\gamma=\sigma=\emptyset$. Thus we take 
    $\bar{\gamma}=\bar{\sigma}=\emptyset$ and 
     we must show next     $\bar{\sigma}_3(\alpha)=\bar{\sigma}_2(\beta)$. 
    Notice that $\beta$  must be a  variable, say $y_2\in \mycal{X}_2$. 
    Since $\set{y_1 \mapsto \sigma_3(\alpha), y_2 \mapsto \sigma_2(\beta)}$ 
    (resp. $\set{y_1  \mapsto \bar{\sigma}_3(\alpha), y_2  \mapsto \bar{\sigma}_2(\beta)}$) appears 
    in the position $\wp_n$ (resp. $\bar{\wp}_n$), 
and $\sigma_3(\alpha)=\sigma_2(\beta)$ then  $\bar{\sigma}_3(\alpha)= \bar{\sigma}_2(\beta)$, 
     since $\bar{\wp}_n \synch \wp_n$.
  \item If $\sigma_2(\beta)$ is a variable, say $y_2\in \mycal{X}_2$, then $\bar{\sigma}_2(\beta)=\sigma_2(\beta)=\beta=y_2$, since  
   $\bar{\sigma}_2 \synch \sigma_2$.     In this case   we have $\gamma=\set{y_2 \mapsto \sigma_3(\alpha)}$ and $\sigma=\emptyset$.
   Thus we take $\bar{\gamma}=\set{y_2 \mapsto \bar{\sigma}_3(\alpha)}$.
     And the invariant (\ref{invariant-sim}) is maintained.
  \end{itemize}
\item If $\sigma_3(\alpha)$ is a variable, say $x_1\in \mycal{X}_1$, then $\bar{\sigma}_3(\alpha)=\sigma_3(\alpha)=\alpha=x_1$. 
  We distinguish two cases  
  depending on the nature of $\sigma_2(\beta)$. 
  \begin{itemize}
  \item If $\sigma_2(\beta)$ is a letter  then $\bar{\sigma}_2(\beta)$ is a letter as well since 
    $\bar{\sigma}_2 \synch \sigma_2$. 
  In this case  $\gamma=\emptyset$     and $\sigma=\set{x_1 \mapsto \sigma_2(\beta)}$.  Therefore we take
    $\bar{\gamma}=\emptyset$ and $\bar{\sigma}=\set{x_1  \mapsto \bar{\sigma}_2(\beta)}$.
  \item  If $\sigma_2(\beta)$ is a variable, say $y_2\in \mycal{X}_2$, then  
     $\sigma_2(\beta)=\bar{\sigma}_2(\beta)=\beta=y_2$ since $\bar{\sigma}_2 \synch \sigma_2$.
   Assume that ${\gamma}=\set{y_2 \mapsto y_0}$, where $y_0 \in \Sigma$ is a letter. In this case 
     we take $\bar{\gamma}=\set{y_2 \mapsto \bar{y}_0}$, where the  choice   of the letter 
  $\bar{y}_0\in C_0$  depends on $y_0$.
  \begin{itemize}
  \item If $y_0  \in \Sigma_{\mycal{A}_1}\cup \Sigma_{\mycal{A}_2}$  then we let $\bar{y}_0:=y_0$.
  \item If $y_0 \in codom(\sigma_1 \uplus \sigma_2)  \setminus \big(\Sigma_{\mycal{A}_1}\cup \Sigma_{\mycal{A}_2}\big)$ then  
    there must exist a variable $z\in \mycal{X}_1\cup \mycal{X}_2$ 
   and a letter $z_0\in C_0$ such that $(z \mapsto y_0) \in \sigma_1 \uplus \sigma_2$ and 
      $(z \mapsto z_0) \in (\bar{\sigma}_1 \uplus \bar{\sigma}_2)$.       We  let  $ \bar{y}_0:=z_0$.
 \item Otherwise, i.e. $y_0$ is a fresh letter that does not appear in $\wp_n$, 
  then  $\bar{y}_0$  must be a fresh letter  as well. 
Since 
\begin{align*}
|codom(\bar{\sigma}_1 \uplus \bar{\sigma}_2)| \le |\mathcal{X}_1|+ |\mycal{X}_2|-1 < |C_0 \setminus (\Sigma_{\mycal{A}_1} \cup \Sigma_{\mycal{A}_2})|
\end{align*}
then 
$  codom(\bar{\sigma}_1 \uplus \bar{\sigma}_2) \subsetneq C_0 \setminus (\Sigma_{\mycal{A}_1} \cup \Sigma_{\mycal{A}_2})$.
Hence we take $\bar{y}_0$ as an arbitrary  
element of the non empty  set 
\begin{align*}
  C_0 \setminus (\Sigma_{\mycal{A}_1} \cup \Sigma_{\mycal{A}_2} \cup codom(\bar{\sigma}_1 \uplus \bar{\sigma}_2))
\end{align*}
  \end{itemize}
\end{itemize}
\end{itemize}
\end{caseinproof}
\end{proofbycases} 
The proof of the direction ''$\Leftarrow$''   is dual  w.r.t. the proof of the direction
 ''$\Rightarrow$''. That is,  it can be obtained by replacing \Eloise by \Abelard, 
  and \Abelard by \Eloise and keeping the same instantiation strategy and the 
 definition of the $\synch$-coherence.
 This ends the proof of the Proposition.\qed
\end{proof}

\section{Appendix for Section \ref{service:synthesis}}
\label{extensions:FVA:appendix}
We extend CFVAs so that 
the transitions are labeled with arbitrary terms over a first-order signature, 
besides the communication symbols indeed. This extended model is called ECFVA.

Let $\mathcal{X}$ be a finite set
of variables, $\Sigma$ a  set of function symbols.
Let $\mathcal{T}(\Sigma,\mathcal{X})$
denote the set of terms built out of the symbols in $\Sigma$ and the
variables in $\mathcal{X}$. We shall denote by $\mterms$ the set 
$\set{!,?}\times \terms$, where $\set{!,?}\cap (\Sigma \cup \mathcal{X})=\emptyset$.
If $t\in \terms$ then $!t$ (resp. $?t$) denotes sending (receiving) the message
$t$. 
A matching problem of a term $t$ by a term $u$, denoted by $\matchq{t}{u}$, 
is solvable iff there is  a substitution $\sigma$ such that $\sigma(t)=u$.
The set of solutions of $\matchq{t}{u}$ is denoted by $\match{t}{u}$. 

The definition of ECFVAs follows. 
\begin{definition}
  A \emph{ECFVA} is a tuple
  $A=\model{\Sigma,\mathcal{X},Q,Q_0,\delta,F,\kappa}$ where $\Sigma$
  is a denumerable set of functional symbols,
  $\mathcal{X}$ is a finite set of variables,
  $Q$ is a finite  set of states, $Q_0\subseteq Q$ is a
  set of initial states, $\delta=Q \times \mterms \to 2^{Q}$ is a
  transition function, $F\subseteq Q$ is a set of accepting states,
  and $\kappa: \mycal{X} \rightarrow  2^Q$ is the refreshing function that associates 
  to every  variable the (possibly empty) set of states where it is refreshed.
\end{definition}

We  define  the  \emph{mirror} of a word $\omega=
\frac ?! l_1\cdot \frac ?! l_n\cdot \ldots$ as the word
$\dual\omega = \frac !? l_1\cdot \frac !? l_n\cdot \ldots$.  
The definition of configuration and run for ECFVAs follows. 
\begin{definition}
\label{run:lang:FVA:def}
Let   $\mycal{A}=\model{\Sigma,\mathcal{X},Q,Q_0,\delta,F,\kappa}$ be a ECFVA.  
A \emph{configuration}  is a pair  $(q,M)$ where $q \in Q$ and 
$M:\mycal{X} \pfun  \Sigma$  is a partial function.
 We define a transition relation  over the  configurations 
  as follows:  $(q_1,M_1)\uberTo{u} (q_2,M_2)$, where $u\in \overline{\mycal{T}}(\Sigma)$,
   iff there exist a term $t \in \mterms$, 
  such that  $q_2\in \delta(q_1,t)$, and a substitution $\sigma=(\match{M_1(t)}{\dual{u}})$
  so that $M_2= (M_1 \uplus \sigma)_{| D}$, where $D=dom(M_1 \uplus \sigma) \setminus  \kappa^{-1}(q_2)$.
   A finite word $u=u_1u_2\ldots u_n \in {\overline{\mycal{T}}(\Sigma)}^*$ is \emph{recognized} by 
$\mycal{A}$ iff  there exists a run
$(q_0,M_{0}) \uberTo{u_1}(q_1,M_1)\uberTo{u_2} \ldots \uberTo{u_n}(q_n,M_n)$, such that 
$M_{0}=\emptyset$, $q_0\in Q_0$ and $q_n \in F$.
 The set of words recognized by $\mycal{A}$ is denoted by
$L(\mycal{A})$.
\end{definition}

\begin{definition}
The  asynchronous product $\otimes$ of $n$ ECFVAs 
$\mathcal{A}_i=\model{\Sigma_i,\mathcal{X}_i,Q_i,Q^i_0,\delta_i,F_i,\kappa_i}$
 is\footnote{Up to variable  renaming, 
 we assume that $\mycal{X}_i\cap \mycal{X}_j=\emptyset$, for all $i\neq j$.} an ECFVA: $\mathcal{A}_1 \otimes \cdots \otimes \mathcal{A}_n=\model{\Sigma,\mathcal{X},Q,Q_0,\delta,F,\kappa}$, 
where:     
\begin{description}             
\item[$\bullet$] $\Sigma=\cup_{i=1,\ldots,n}\Sigma_i$, 
\item[$\bullet$]$\mathcal{X}=\cup_{i=1        ,\ldots,n}\mathcal{X}_i$, 
\item[$\bullet$] $Q=Q_1\times \cdots \times Q_n$,
\item[$\bullet$]  $Q_0=Q^1_0 \times \cdots \times Q_0^n$, $F=F_1\times \cdots \times F_n$, 
 \item[$\bullet$]  $\delta$ is defined by:  
 $\boldsymbol{q} \in \delta(\boldsymbol{p},t)$ iff for some $i$, $\pi_i(\boldsymbol{q}) \in \delta_i(\pi_i(\boldsymbol{p}),t)$, and for 
 all $j\neq i$ we have that $\pi_j(\boldsymbol{q})=\pi_j(\boldsymbol{p})$, where $\pi_i$ denotes the  projection
 along the  $i^{th}$-component, and
 \item[$\bullet$] $\kappa$ is defined by:  
  $ \boldsymbol{p} \in \kappa(x) $ iff for some $i$, $\pi_i(\boldsymbol{p}) \in \kappa_i(x)$.
\end{description}
\end{definition}

\subsection{Undecidability of the \gsim{simulation} problem for ECFVAs}
\label{annex:undecid:gsim:sec}
\begin{theorem}
The \gsim{simulation} is undecidable for ECFVAs in which the labels  are 
terms over a signature containing a unary symbol.    
\end{theorem}

We reduce the halting problem of 2 counter machines to the simulation problem 
for ECFVAs. 
Let us consider a  deterministic 2-counter machine $M$ with set of states $\cal Q$ and 
such that $q_0$ is the initial state and $q_f$ the final one (from where no transition is possible). 
A configuration of the machine can be represented by a term $q(s^n(0),s^m(0))$ 
where $q$ is the state, and $n$ (resp. $m$) the value of the first (resp. second) counter.
The initial configuration of $M$ is $q_0(s^i(0),s^j(0))$
We encode every transition $l : q(u,v)\rightarrow q'(u',v')$ of the machine by 
a (deterministic) ECFVAs $A_l$ as follows (we consider 
only the cases when the first counter is incremented, decremented or tested, the  cases for the second counter are analogous):
$\Sigma_l = \{q, q'\} \cup \{s , 0\}$, $\mathcal{X}_l$ is a finite set of variables, 
$Q_l= \{p^0_l,p^1_l,p^2_l,p^3_l,p^4_l,p^5_l\}$  
and the set of transitions $\delta_l$ (where $u,v \in \mathcal{X}_l$ ) : \\
\begin{center}
\begin{tabular}{lcll}
\multicolumn{3}{c}{Instruction $l$} & \multicolumn{1}{c}{Set of transitions $\delta_l$}  \\
$q(u,v)$ & $\rightarrow$ & $q'(s(u),v)$ & $\{p^0_l\xrightarrow{?q}p^1_l, p^1_l \xrightarrow{?u}p^2_l, p^2_l\xrightarrow{?v}p^3_l,p^3_l\xrightarrow{!q'}p^4_l,p^4_l\xrightarrow{!s(u)}p^5_l,p^5_l\xrightarrow{!v}p^0_l\}$\\
$q(s(u),v) $ &$\rightarrow$ &$ q'(u,v)$ & $\{p^0_l\xrightarrow{?q}p^1_l, p^1_l\xrightarrow{?s(u)}p^2_l, p^2_l\xrightarrow{?v}p^3_l,p^3_l\xrightarrow{!q'}p^4_l,p^4_l\xrightarrow{!u}p^5_l,p^5_l\xrightarrow{!v}p^0_l\}$\\
$q(0,v) $ & $\rightarrow $ & $q'(u,v)$ & $\{p^0_l\xrightarrow{?q}p^1_l, p^1_l\xrightarrow{?0}p^2_l, p^2_l\xrightarrow{?v}p^3_l,p^3_l\xrightarrow{!q'}p^4_l,p^4_l\xrightarrow{!0}p^5_l,p^5_l\xrightarrow{!v}p^0_l\}$
\end{tabular}\\
\end{center}

Now we build a client automata $C_M$ such that $\Sigma =Q  \cup \{s , 0\}$, $\mathcal{X}$ 
is a finite set of variables, the set of states is $Q_M=\{I,F, c^0,c^1,c^2,c^3,c^4,c^5\}$, 
$I$ is the unique initial state and all states are accepting.

The set of transitions of $C$ is the union of the following ones (where $u,v \in \mathcal{X}$)  :\\
\begin{center}
\begin{tabular}{ll}
Initial sequence:& $\{I\xrightarrow{!p^0} I',  I'\xrightarrow{!s^i(0)} I" ,  I" \xrightarrow{!s^j(0)} c^0\}  $\\
For all $q\in\cal Q$:& $\{c^0\xrightarrow{?q}c^1, c^1 \xrightarrow{?u}c^2, c^2\xrightarrow{?v}c^3,c^3\xrightarrow{!q}c^4,c^4\xrightarrow{!u}c^5,c^5\xrightarrow{!v}c^0\} $\\
Final loop:& $\{c^0\xrightarrow{?q_f} F, F \xrightarrow{!q_f } F \}$
\end{tabular}\\
\end{center}
The Client automata starts by sending the initial configuration of $M$, 
then she  simply sends back the  configurations she  receives 
till she receives $q_f$ the final state of $M$. If this happens $C_M$ 
enters a loops by keep on sending  back $q_f$. 
Since no transitions from  $q_f$ exists in $M$ there is no service automaton 
that can accept the message $q_f$. Hence the 
2-counter automata halts iff $C_M$ cannot be simulated by the asynchronous product of 
automata $A_l$.


\section{Further results on FVAs}
\label{more}
{For convincing the reader, we present here further results which have 
not been presented in the core of the paper.}

We  provide a fine  comparison between FVAs and 
 $n$-FVAs, then we define deterministic FVAs and study some of their 
properties.  
\subsection{The  $n$-FVAs and their expressiveness}
\label{NFVA:sec:appendix}

To compare $n$-FVAs  and FVAs, 
 the definition of the relation of simulation and bisimulation for FVAs is needed.

\begin{definition}
\label{sim:FVA:def}
Let   $\mycal{A}_1=\model{\Sigma,\mathcal{X}_1,Q_1,q^{1}_0,\delta_1,F_1,\kappa_1}$
 and  $\mycal{A}_2=\model{\Sigma,\mathcal{X}_2,Q_2,q^2_0,\delta_2,F_2,\kappa_2}$ be two FVAs where
 $\mycal{X}_1 \cap \mycal{X}_2=\emptyset$.
A simulation of  $\mycal{A}_1$ by $\mycal{A}_2$ is a relation 
$\unlhd  \subseteq (\zeta \times Q_1) \times (\zeta \times Q_2)$ such that
\begin{itemize}
\item if $(\sigma_1,q_1) \unlhd (\sigma_2,q_2)$ and if $q'_1 \in \delta_1(q_1,x_1)$ for a variable 
 $x_1 \in \mycal{X}_1$,  and   $\gamma_1:\Eu{V}(\sigma_1(x_1)) \uberTo{} \Sigma$ is a substitution and
\begin{align*}
    (\sigma_1,q_1) \uberTo{\gamma_1(\sigma_1(x_1))} (\underbrace{\sigma_1\cup \set{(x_1,a)}_{|D_1}}_{\sigma'_1},q'_1),
\end{align*}  
 where $D_1=dom(\sigma_1)\setminus \kappa_1^{-1}(q'_1)$, 
 then there exist a variable $x_2 \in \mycal{X}_2$ and a transition 
  $q'_2 \in \delta_2(q_2,x_2)$ and a substitution $\gamma_2:\Eu{V}(\sigma_2(x_2)) \uberTo{}\Sigma$ 
such that $\sigma_1(x_1)=\sigma_2(x_2)$ and 
 \begin{align*}
   (\sigma_2,q_2) \uberTo{ \gamma_2(\sigma_2(x_2))} (\underbrace{\sigma_2\cup \set{(x_2,a)}_{|D_2}}_{\sigma'_2}, q'_2)
 \end{align*}
 where   $(\sigma'_1,q'_1) \unlhd (\sigma'_2,q'_2)$ and  $D_2=dom(\sigma_2)\setminus \kappa_2^{-1}(q'_2)$.
\item The cases  when $\mycal{A}_1$ performs a transition labeled by a  letter   
and  $\mycal{A}_2$ 
   replies  by a  transition labeled by either  a  letter or a free variable  are handled in the usual way. 
\item $(\emptyset,q^1_0) \unlhd (\emptyset,q^2_0)$.
\item If $(\sigma_1,q_1) \unlhd (\sigma_2, q_2)$ with $q_1\in F_1$ then $q_2\in F_2$.
\end{itemize}
\end{definition}

\begin{lemma}
\label{sim:prop:lem}
The simulation relation $\unlhd$  of FVAs enjoys the following properties:
\begin{enumerate}
\item it is a preorder, i.e. reflexive and transitive, 
\item  it  implies language inclusion, 
i.e. if $\mycal{A} \unlhd \mycal{B}$ then $L(\mycal{A}) \subseteq L(\mycal{B})$,  
 for two FVAs $\mycal{A}$ and $\mycal{B}$, and
\item it is decidable.
\end{enumerate}
\end{lemma}
\begin{proof}
Items \emph{1} and \emph{2} are immediate. 
For the Item \emph{3}, the same technique used in the proof that the \gsim{simulation} is decidable 
(Theorem \ref{main:decidable:th}) can be 
 reused: there is a finite set $C$ of letters  such that there is a simulation where the variables are instantiated 
from the infinite set $\Sigma$ iff there is a simulation where 
the variables are instantiated from  $C$.
\qed
\end{proof}

The relation of bisimulation for FVAs, denoted hereby $\approx$, can be defined
in the same fashion as  the relation of simulation.

Although   $n$-FVAs  and FVAs recognize the same languages, 
 $n$-FVAs are stronger than $(n-1)$-FVAs  in the  following sense: 

\begin{theorem}
\label{vaut:complement:prop}
For every $n \ge 2$,  there is an   $n$-FVA $\mycal{H}_n$ 
so  that there is no $n'$-FVA $\mycal{H}_{n'}$  such  that   $\mycal{H}_n$ and
$\mycal{H}_{n'}$ are bisimilar and $n'<n$.
\end{theorem}
\begin{proof}
Let $\mycal{H}_n=\model{\Sigma,\mathcal{X},Q,q_0,\delta,F,\kappa}$  
 be the $n$-FVA  depicted below  and  defined by 
\begin{align*}
&\mycal{X}=\set{x_1,\ldots,x_n}, \\
&Q = \set{q_{-1},q_0,\ldots,q_n} \cup  \set{q_{i}^1, i=1,\ldots,n} \cup \set{q_{i}^2, i=1,\ldots,n},\\
&F= Q \\
  & \delta =  \set{q_{-1}\uberTo{x_1}q_0} \cup \set{q_i\uberTo{x_{i+1}} q_{i+1}, i=0,\ldots,n-1} \cup
     \set{q_i\uberTo{x_1,\ldots,x_{i+1}} q_{i}^1, i=1,\ldots,n-1} \cup  \\
   &\quad \quad \set{q_{i}^1\uberTo{b} q_i^2, i=2,\ldots,n-1} \\
& dom(\kappa)= \emptyset,
\end{align*}
where $b\in \Sigma$.
 We show that 
there is no $(n-1)$-FVA  $\mycal{B}_{n-1}$ that $\mycal{H}_n \approx \mycal{B}_{n-1}$.
Towards a contradiction: assume the existence of such  
 $ \mycal{B}_{n-1}=\model{\Sigma,\mathcal{X}',Q',Q'_0,\delta',F',\kappa'}$.
There exist two substitutions 
 $\sigma_{n-1}:\mycal{X}\uberTo{} \Sigma$ and 
 $\sigma'_{n-1}:\mycal{X}' \uberTo{} \Sigma$,  
 and a state $q'_{n-1} \in Q'$
such that $(\sigma_{n-1},q_{n-1}) \approx (\sigma'_{n-1},q'_{n-1})$.
Notice that $dom(\sigma_{n-1})=\set{x_1,\ldots,x_{n-1}}$. We argue next
that the transition $q_{n-1} \uberTo{(x_1,\ldots,x_n)} q_{n-1,1}$ of $\mycal{H}_n$ can not be simulated 
by any transition of $\mycal{B}_{n-1}$ outgoing from $q'_{n-1}$.
Each   transitions outgoing from $q'_{n-1}$ is labeled  by a letter or an 
$(n-1)$-labels of variables   $(x'_1,\ldots,x'_{n-1})$. Notice that 
  when there exist  $i,j$ such that 
if $\sigma'(x'_i) \neq \sigma'(x'_j)$, then 
 one of the  outgoing   transitions  from $q'_{n-1}$  is possible, 
but this transition must be matched by the transition $q_{n-1}\uberTo{x_n} q_n$ in $\mycal{H}_{n}$. 
And the $b$-transition of $\mycal{B}_{n-1}$ can not be matched by any transition in $\mycal{H}_n$
since there is no outgoing transition from $q_n$.

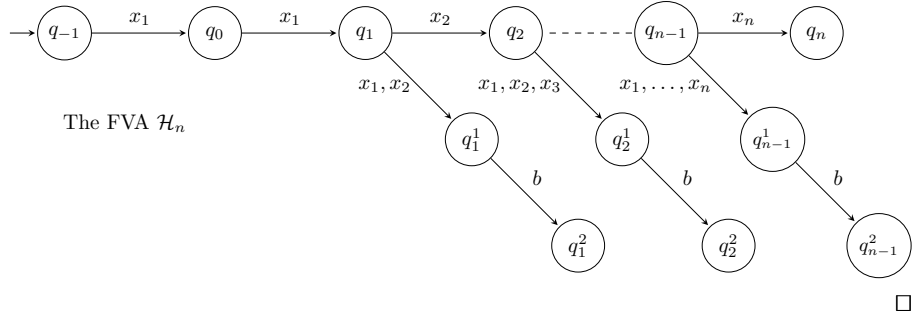
\begin{center} 
\scalebox{0.8}
{\begin{tikzpicture}[shorten >=1pt,node distance=3.5cm, bend angle=60,
   on grid,auto, initial text=, >=stealth] 
  \begin{scope}[xshift=0cm, yshift=0cm,node distance=2.5cm]
   \node[state,initial] (q_0)   {$q_{-1}$}; 
   \node[state] (q_{0p}) [right =of q_0] {$q_0$}; 
   \node[state] (q_1) [right =of q_{0p}] {$q_1$}; 
   \node[state] (q_2) [right =of q_1] {$q_2$}; 
   \node[state] (q_3) [right =of q_2] {$q_{n-1}$}; 
   \node[state] (q_4) [right =of q_3] {$q_{n}$}; 

   \node[state] (q_{11}) [below  right =of q_1] {$q_1^1$}; 
   \node[state] (q_{12}) [below  right =of q_{11}] {$q_1^2$}; 

   \node[state] (q_{21}) [below right =of q_2] {$q_2^1$}; 
   \node[state] (q_{22}) [below right =of q_{21}] {$q_2^2$}; 

   \node[state] (q_{31}) [below right =of q_3] {\begin{small}$q_{n-1}^1$\end{small}}; 
   \node[state] (q_{32}) [below right =of q_{31}] {\begin{small}$q_{n-1}^2$\end{small}}; 

    \path[->] 
    (q_0)  edge   node  {$x_1$} (q_{0p})
    (q_{0p})  edge   node  {$x_1$} (q_1)
    (q_1)  edge  node   {$x_2$} (q_2)
    (q_3)  edge   node  {$x_n$} (q_4)

    (q_1)  edge   node [left]  {$x_1,x_2$} (q_{11})
    (q_{11})  edge   node    {$b$} (q_{12})

    (q_2)   edge   node [left] {$x_1,x_2,x_3$} (q_{21})
    (q_{21})  edge   node  {$b$} (q_{22})

    (q_3)  edge   node [left] {$x_1,\ldots,x_n$} (q_{31})
    (q_{31})  edge   node  {$b$} (q_{32})

    -- (1,-1.5) node {The FVA $\mycal{H}_n$};
\draw[dashed,rounded corners=8pt] (8.06,0) -- (9.5,0);
 \end{scope}
\end{tikzpicture}}
\end{center}
 \qed
\end{proof}


\subsection{Deterministic FVAs.}
 \label{DFVA:sec:appendix}
We  define  \emph{deterministic} FVAs, 
(DFVAs, for short) in terms of runs. Then we give a  syntactic 
characterization of them.

\begin{definition}
A FVA $\mathcal{A}$ is deterministic if  for every word 
 $w \in \Sigma^{\star}$ there exists  at most one run  of  $\mathcal{A}$ on $w$. 
\end{definition}

\begin{theorem}
\label{synt:charat:DFVA:th}
Let $\mycal{A}$ be a FVA. 
Then $\mycal{A}$ is not deterministic iff there exists an accessible state $q$ 
with two outgoing transitions satisfying one of the conditions:
\begin{enumerate}
\item the transitions are labeled with the same letter; 
\item one of the transitions is labeled by a variable.
\end{enumerate} 
\end{theorem}

It is clear that the above conditions are sufficient and necessary.

\begin{proposition}
There is a FVA $\mycal{A}$ such that  no  DFVA $\mycal{D}$ satisfies $L(\mycal{A})=L(\mycal{D})$.
\end{proposition}
\begin{proof}
Let $a,b$ be two letters  in $\Sigma$, and let  $\mycal{L}=\set{z \gvert z\in \Sigma} \cup \set{ab}$.
Indeed the language $L$ is FVA-recognizable. Towards a contradiction: assume the existence 
 of a DFVA $\mycal{D}$ such that $L(\mycal{D})=\mycal{L}$. Let $q_0$ be the initial state of $\mycal{D}$.
 By following the syntactic 
characterization of DFVAs given in Theorem \ref{synt:charat:DFVA:th}, 
we have that either \emph{(i.)} all the transitions outgoing from $q_0$ 
 are labeled with letters, 
and in this case the language $\set{z \gvert z \in \Sigma}$ can not be recognized by 
 $\mycal{D}$ since $\Sigma$ is infinite, which is a contradiction, or
\emph{(ii.)} there is only one transition  outgoing from $q_0$  and labeled with a variable.
 Let  $q_0\uberTo{x}q_1$ be such  transition. 
In this case, there must be a transition  $q_1\uberTo{b}q_f$ in $\mycal{D}$ where $q_f$ is a final 
state.
This means that the set of words $az$, where $z \in \Sigma$ is recognized by $\mycal{D}$.
This is a contradiction.   \qed
\end{proof}

\begin{corollary}
Deciding if a FVA is deterministic is NL-Complete. 
\end{corollary}
\begin{proof}
The upper bound follows from the fact that we can guess a condition and check whether it  is 
violated. On the other hand, NL is closed under complementation. 
The lower bound follows from a standard reduction from the reachability for digraphs. 
\qed
\end{proof}

\begin{proposition}
For DFVAs, the membership and the universality problems  are in
PTIME. 
\end{proposition}
\begin{proof}
We only discuss the complexity of the universality since the 
membership problem is straightforward.  
Let $\mycal{A}$ be a DFVA.
 Recall that 
to check whether $\mycal{A}$ 
 is universal we first construct an equivalent 
 FVA $\mycal{A}'$
 in which all the transitions are labeled with free variables, see the proof of Thereom \ref{univ:FVA:th}.
To construct $A'$ one may first eliminate all the transitions 
of $\mycal{A}$ labeled with letters. This yields a DVFA $\mycal{A}^{c}$ whose
 structure  is a tail-cycle in which 
 all the transitions are labeled with variables. 
Hence, the universality of $\mycal{A}^c$ can be done in polynomial time.
  \qed
\end{proof}

\begin{proposition}
\label{contain:FVA:DBFAV:Th}
The containment problem  $L(\mycal{A}) \subseteq L(\mycal{D})$  for two FVAs $\mycal{A}$ and $\mycal{D}$ 
where $\mycal{D}$ is deterministic, is  decidable.
\end{proposition}
\begin{proof}
  We shall show that  $L(\mycal{A})  \subseteq L(\mycal{D})$ 
iff  $\mycal{D}$  simulates $\mycal{A}$. 
The direction  $(\Rightarrow)$ has been proven in item \emph{2} of Lemma \ref{sim:prop:lem}. 
The direction $(\Leftarrow)$ follows from the fact that every accepting run by $\mycal{A}$ over a 
word $w \in \Sigma^{\star}$ can be simulated by a \emph{unique} run by $\mycal{D}$ over $w$.
\end{proof}





\end{document}